\documentclass[%
 reprint,
 superscriptaddress,
nofootinbib,
 amsmath,amssymb,
 aps,
]{revtex4-2}

\usepackage{subcaption}
\usepackage{pgfplots} 
\usepackage{graphicx}
\usepackage{dcolumn}
\usepackage{bm}
\usepackage{siunitx}
\usepackage{rotating}
\usepackage{float} 
\restylefloat{table}

\usepackage{siunitx}

\usepackage{soul}

\pgfplotsset{compat=1.14}

\begin{document}

\title{Photon statistics and signal to noise ratio for incoherent diffraction imaging}

\author{Fabian Trost}
 \email{fabian.trost@cfel.de}
 \affiliation{Center for Free-Electron Laser Science, Deutsches Elektronen Synchrotron DESY, Notkestra{\ss}e 85, 22607 Hamburg, Germany}
\author{Kartik Ayyer}%
 \affiliation{Max Planck Institute for the Structure and Dynamics of Matter, Luruper Chaussee 149, 22761 Hamburg, Germany}
 \affiliation{Center for Free-Electron Laser Science, Luruper Chaussee 149, 22761 Hamburg, Germany}
 \affiliation{The Hamburg Center for Ultrafast Imaging, Universit{\"a}t Hamburg, Luruper Chaussee 149, 22761 Hamburg, Germany}
\author{Henry Chapman}
 \affiliation{Center for Free-Electron Laser Science, Deutsches Elektronen Synchrotron DESY, Notkestra{\ss}e 85, 22607 Hamburg, Germany}
 \affiliation{The Hamburg Center for Ultrafast Imaging, Universit{\"a}t Hamburg, Luruper Chaussee 149, 22761 Hamburg, Germany}
 \affiliation{Department of Physics, Universit{\"a}t Hamburg, Luruper Chaussee 149, Hamburg, Germany}
 \affiliation{Molecular and Condensed Matter Physics, Department of Physics and Astronomy, Uppsala University, Box 516, SE-751 20 Uppsala, Sweden}

\date{\today}

\begin{abstract}
Intensity interferometry is a well known method in astronomy. Recently, a related method called incoherent diffractive imaging (IDI) was proposed to apply intensity correlations of x-ray fluorescence radiation to determine the 3D arrangement of the emitting atoms in a sample. Here we discuss inherent sources of noise affecting IDI and derive a model to estimate the dependence of the signal to noise ratio (SNR) on the photon counts per pixel, the temporal coherence (or number of modes), and the shape of the imaged object.
Simulations in two- and three-dimensions have been performed to validate the predictions of the model. We find that contrary to coherent imaging methods, higher intensities and higher detected counts do not always correspond to a larger SNR. Also, larger and more complex objects generally yield a poorer SNR despite the higher measured counts. The framework developed here should be a valuable guide to future experimental design.

\end{abstract}

\maketitle

\section{\label{sec:intro}Introduction}
The scattering of a spatially and temporally coherent beam from an object gives rise to a far-field diffraction pattern consisting of constructive and destructive interference that encodes that object's structure, an effect that is utilised to obtain atomic-resolution images of the electron density of crystals with x-rays, for example.  If measured in a similar way, the far-field pattern of light emitted by a luminous object, however, appears unstructured since the individual emitters of that object are mutually incoherent. Unlike in the case of coherent diffraction, the phase relationships of spherical waves emanating from elements of the object do not remain constant, and thus over the course of an exposure the measurement averages to the sum of the integrated intensities of those emitters.  If, on the other hand, the far-field pattern is measured with an exposure time shorter than the coherence time of the light, we would indeed observe interference in the form of a speckle pattern~\cite{PhysRevLett.119.053401,Goodman:76}.  Although this speckle pattern would change each time it is measured due to random fluctuations of the phases of the emitters, the integrated intensities nevertheless retain correlations.  This is the basis for intensity interferometry of Hanbury Brown and Twiss, in which the signals measured in independent detectors are correlated. The method was first used to measure the correlation length for radio and visible stars to deduce their diameters~\cite{HBT:1956:StellarInterf}. 

Classen \emph{et al.}~\cite{PhysRevLett.119.053401} proposed to use intensity interferometry of x-ray fluorescence to reconstruct the three-dimensional arrangement of a particular species of atom in a sample such as a protein crystal, a method referred to as incoherent diffractive imaging (IDI).  Fluorescence is generated by the transition of a valence electron into the core hole created by x-ray photoionisation.  For transition metal elements such as Fe or Mn, the cross section for photoabsorption exceeds that of coherent scattering by a factor of about 250~\cite{NIST}, producing about 50 K$\alpha$ fluorescence photons per coherently scattered photon. The wavelength of the emission is on the order of \SI{1}{\angstrom} and the lifetime---and thus the coherence time $\tau_\text{c}$---is of the order of \SI{0.4}{\fs}.  
It was suggested that the femtosecond-duration pulses produced by x-ray free-electron lasers would generate fluorescence from a sample within a burst that could then be measured with an integrating detector to compute the intensity-intensity correlation, as recently demonstrated by Inoue \emph{et al.}~\cite{Inoue:gb5094}.  As yet, only the width of the x-ray spot focused onto a fluorescing metal foil has been determined by this method---the image of a more complicated structure such as a crystal is yet to be demonstrated.  

The design of IDI experiments requires an analysis of the signal to noise ratio (SNR) that can be achieved by the method and how this depends on various experimental parameters. Estimates of the achievable SNR in images obtained by intensity interferometry of general scenes have been presented in the context of astronomy~\cite{HBT:1956:StellarInterf,HBT_1957,HBT_1957_II,Foellmi:2009,Nunez:2012,Strekalov:2013,Rou:2013,Fried:14,Gamo:66}. These studies suggest the SNR scales with intensity---that is, with the number of photons measured per coherence mode---and should improve with the square root of the number of detector pairs (and hence correlations) that contribute to the measurement. However, prior works omit considerations of the consequences of performing pair-correlations of intensities measured simultaneously on many independent detectors (or detector pixels) as will be the case when fluorescing atoms are stimulated by a femtosecond-duration x-ray pulse. This, as we find here, has a profound influence on the achievable SNR.  

The situation for IDI can be compared with coherent diffractive imaging (CDI) based on elastic scattering, where for a well-designed experiment the noise in the measured integrated intensities is dominated by the Poisson statistics of the photons and so higher measured counts yields a higher SNR. Although this contention also holds for IDI, the situation is more complicated because of the way the signal is constructed from a correlation of intensities. It has not been readily obvious what other factors the SNR depends upon, and what conditions must be met for a feasible experiment. Our analysis is based upon a classical (wave optics) approach, combined with photon statistics, to determine the statistics of detected signals and the corresponding statistics of their correlations. 
After briefly reviewing the method of IDI in Sec.~\ref{sec:IDI}, compared with CDI, we introduce the statistics of the correlation function in Sec.~\ref{sec:noise}. These are then used to estimate the relative SNRs in Sec.~\ref{sec_SNR} as a function of experimental parameters and the object shape, which we compare with numerical simulations. We assess the feasibility of imaging different types of structures using snapshot x-ray fluorescence measurements in Sec.~\ref{sec:summary} as well as the imaging of stars at high angular resolution using arrays of visible telescopes.  Our results show that the complexity of the structure is a crucial factor in the ability to determine the first-order coherence function $g^{(1)}$ of the light-field (equal to the normalized Fourier transform of the spatial distribution of emitters) from measurements of the second-order coherence function $g^{(2)}$ (the normalized intensity autocorrelation).

\section{Incoherent Diffractive Imaging}
\label{sec:IDI}
\begin{figure}
   \def\svgwidth{0.60\linewidth}
   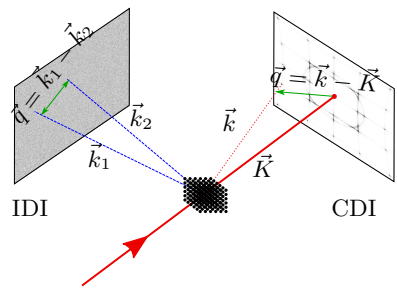
\caption{\label{fig_IDI_CDI_DEMO} Schematic sketch of an IDI setup to demonstrate the differences from a coherent diffractive imaging (CDI) setup. Since fluorescence is emitted isotropically the IDI detector placement does not depend on the incident beam (but can be placed, where coherent diffraction is suppressed by polarization). In IDI, structural information is obtained by correlation, therefore the distance between two pixels gives rise to a certain $\vec{q}$ while in CDI $\vec{q}$ is defined relative to the incident beam.}
\end{figure}
Figure~\ref{fig_IDI_CDI_DEMO} depicts a general scattering experiment that gives access to both a ``coherent diffraction imaging" (CDI) measurement and an IDI measurement. In CDI, the interference of elastically-scattered waves is recorded as a diffraction pattern, shown here in the forward direction.  For a particular position on the detector in the far field, specified by the direction of the wavevector $\vec{k}$, this interference can be calculated by summing over all rays originating from a source point (assumed here at infinity) and scattering from the elemental scatterers in the sample (e.g. atoms) to arrive at the detector.  The relative phases of these rays depend on their path differences and are given by $(\vec{k}-\vec{K})\cdot \vec{r}$, where $\vec{r}$ is the position of the scatterer relative to some arbitrary origin and $\vec{K}$ is the common wavevector of the rays incident on the sample.  The phases are further modulated by the complex-valued scattering factor $f$ of each scatterer, giving rise to a diffraction pattern $I(\vec{q})=|\sum_i f_i \exp (i \vec{q} \cdot \vec{r_i})|^2$, with $\vec{q} = \vec{k}-\vec{K}$. Since the scattering is elastic, the magnitudes of $\vec{k}$ and $\vec{K}$ are equal and the diffraction amplitudes represent Fourier components $\tilde\rho(\vec{q})$ measured on a spherical manifold of radius $2\pi/\lambda$ (for a wavelength $\lambda$) that passes through the origin $\vec{q}=0$. This manifold is referred to as the Ewald sphere~\cite{cowley1995}.

We can compare CDI to IDI by considering monochromatic fluorescence emitted from the sample and detected on the grey-coloured detector in Fig.~\ref{fig_IDI_CDI_DEMO}.
If measured with an exposure much shorter than the coherence time of the fluorescence, and ignoring for the moment any sources of noise or quantization as well as polarization, the waves originating from the elemental emitters of strength $s_i$ in the sample will interfere in a similar way to those formed by elastic scattering discussed above, as
\begin{equation}
  \label{eq:random-speckle}
  \begin{aligned}
    I(\vec{k_1}) &=\left|\sum_i s_i \,e^{ i(\vec{k_1}\cdot \vec{r}_i+\phi_i)}\right|^2 \\ 
    &=\sum_{ij}s_i \,s^*_j \,e^{ i(\phi_i - \phi_j)} \,e^{ i\vec{k}_1 \cdot
    (\vec{r_i}-\vec{r_j})}
    \end{aligned}
\end{equation}
A difference to CDI is that the phases $\phi_i$ of the waves are random and uncorrelated. The diffraction pattern is thus a speckle pattern, with values that follow a negative exponential distribution~\cite{Goodman:76} and speckles of a width inversely proportional to the width of the object. An example of such a pattern is given in Fig.~\ref{fig_SpeckleDemo}. The different realisations of the random phases $\phi_i$ each time a measurement is made will give rise to a different speckle pattern. At first sight it would seem that structural information would not readily be discernable, since over an ensemble $\{p\}$ of repeated measurements $\langle \exp (i(\phi_{i,p}-\phi_{j,p})) \rangle_p = \delta_{ij}$ so that $\langle I_p(\vec{k}_1) \rangle_p = \sum_i |s_i|^2$ becomes featureless.  For an object consisting of $N_\mathrm{E}$ identical emitters, this is equal to $N_\mathrm{E}\,|s|^2$.   However, if we instead take correlations of integrated intensities at positions $\vec{k}_1$ and $\vec{k}_2$ within an exposure, and then average these correlations over many patterns, then it can be seen that
\begin{equation}
  \label{eq:corr-avg}
  \langle I_p(\vec{k}_1)\,I_p(\vec{k}_2) \rangle_p = N_\mathrm{E}^2 |s|^4 + \left| \sum_j |s|^2 \,
    e^{i(\vec{k}_1-\vec{k}_2)\cdot \vec{r}_j} \right|^2.
\end{equation}
Furthermore, additional averaging can be carried out over all pairs of pixels with a given displacement $\vec{q}=\vec{k}_1-\vec{k}_2$ so that
\begin{equation}
  \label{eq:corr-avg2}
  \langle \langle I_p(\vec{k})\,I_p(\vec{k}-\vec{q}) \rangle_{\vec{k}} \rangle_p = N_\mathrm{E}^2 |s|^4 + \left| \sum_j |s|^2 \,
    e^{i\vec{q}\cdot \vec{r}_j} \right|^2.
\end{equation}
For a pixelated detector that covers a particular solid angle of the emission from the object (such as seen in Fig.~\ref{fig_IDI_CDI_DEMO}), the samples of $\vec{q}$ cover a volume of reciprocal space given by the autocorrelation of the Ewald sphere surface provided by that solid angle~\cite{PhysRevLett.119.053401}.  Therefore, compared with a coherent diffraction pattern that encodes data only on a two-dimensional manifold of reciprocal space, IDI encodes three-dimensional information.

We note that although Eqn.~\ref{eq:corr-avg2} was derived using classical wave optics and the assumption of random phases, this reproduces the derivation based upon quantum mechanics presented in Classen \emph{et al.}~\cite{PhysRevLett.119.053401}. This equation shows that the autocorrelation of the values measured within the coherence time of emitters is equal to the square of the total emitted power added to the square modulus of the Fourier transform of the distribution of emitters $S(\vec{r}) = |s(\vec{r})|^2$, as
\begin{equation}
  \label{eq:G2}
   G^{(2)}(\vec{q}) =\left|\tilde{S}(0) \right|^2 + \left| \tilde{S}(\vec{q}) \right|^2  ,
\end{equation}
where $\tilde{S}(\vec{q})$ is the Fourier transform of $S(\vec{r})$.  Obtaining the emitter structure $S(\vec{r})$ from the square modulus of its Fourier transform is the same phasing problem that faces CDI, and this can be tackled using the same tools, such as iterative projection algorithms~\cite{Elser:03, Elser418}. 

It is convenient to carry out analysis on normalised quantities,  $g^{(1)}(\vec{q}) = \tilde{S}(\vec{q}) / \tilde{S}(0)$, so that Eqn.~\ref{eq:G2} becomes
\begin{equation} \label{eq_g2}
    g^{(2)}_\text{TLS}(\vec{q}) = 1 + \left| g^{(1)}(\vec{q}) \right|^2 \, .
\end{equation}
More generally, for measurements made with only partial temporal coherence, the contrast of the $g^{(1)}$ signal will be reduced by the visibility $\beta$ with $0<\beta<1$ so that 
\begin{equation} \label{eq_g2_beta}
    g^{(2)}_\text{TLS}(\vec{q}) = 1 + \beta\,\left| g^{(1)}(\vec{q}) \right|^2 .
\end{equation}
This is known as the Siegert relation~\cite{PhysRevLett.119.053401,Goodman}.  Thus far, the derivation has been purely classical, and this relationship holds for thermal light source (TLS) emitters, and hence the subscript in Eqns.~\ref{eq_g2} and \ref{eq_g2_beta}. In the case of inner-shell x-ray fluorescence from single atoms, we can not assume a thermal light source. Instead a more accurate model assumes a source composed of single photon emitters (SPEs). In this case, the intensity auto-correlation Eqn.~\ref{eq_g2} must be corrected to account for the inability of an atom to emit another photon within a coherence time~\cite{PhysRevLett.119.053401}:
\begin{equation}\label{eq_g2_SPE}
    g^{(2)}_\text{SPE} = 1 - 2/N_\mathrm{E} + \left| g^{(1)}(\vec{q}) \right|^2 .
\end{equation}
For objects with a large number of emitters, the difference between these expressions for SPEs and TLSes vanishes. Here we restrict the analysis to the TLS case.

\section{Sources of noise}
\label{sec:noise}

\begin{figure}
\begin{tabular}{c c c}
    & $M=1$ & $M=20$ \\
    
    \rotatebox[origin=c]{90}{$\mu = 20$}
    &        
    \begin{subfigure}[c]{0.33\linewidth}
       \resizebox{\linewidth}{!}{\frame{\includegraphics{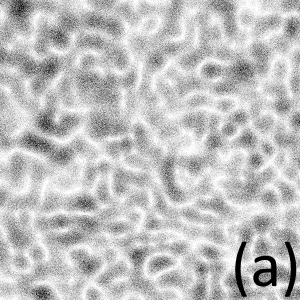}}}
    \end{subfigure}
    &
    \begin{subfigure}[c]{0.33\linewidth}
       \resizebox{\linewidth}{!}{\frame{\includegraphics{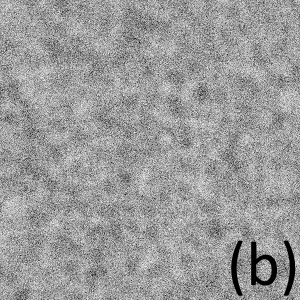}}}
    \end{subfigure}\\ \\[-1.5ex]
    \rotatebox[origin=c]{90}{$\mu = 0.05$}
    &
    \begin{subfigure}[c]{0.33\linewidth}
       \resizebox{\linewidth}{!}{\frame{\includegraphics{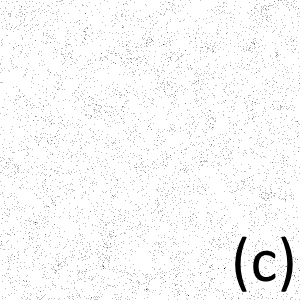}}}
    \end{subfigure}
    & 
    \begin{subfigure}[c]{0.33\linewidth}
       \resizebox{\linewidth}{!}{\frame{\includegraphics{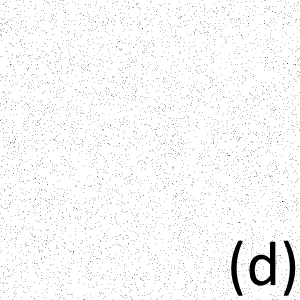}}}
    \end{subfigure}
\end{tabular}
\caption{\label{fig_SpeckleDemo}Simulated speckle patterns of an object of a total of \num{3375} emitters arranged in a simple cubic lattice, for different mean photon count per pixel $\mu$ and temporal coherence modes, $M$. See text and Appendix~\ref{sec_A_3DSim} for more details. }
\end{figure}

As evident from Eqn.~\ref{eq_g2_beta} the determination of $\left| g^{(1)}(\vec{q}) \right|^2$ ideally requires measurements with a high intensity per pixel (sufficient to neglect shot noise), a large number ($N_\mathrm{P}$) of recorded exposures, and full temporal coherence. Such an ideal of course cannot be met, and we examine here the effects on the $g^{(2)}$ signal for non-ideal conditions. We make a distinction between two sources of noise as well as a reduction of contrast.

The first source of noise is that caused by a finite integrated intensity per pixel and the quantum nature of light, commonly known as ``shot noise''. As this noise produces Poissonian statistics we simply refer to it as ``Poisson noise''.
The second source of noise arises due to the finite number of patterns, which we refer to as ``phase noise'' since its origin lies in random phases of the emitted waves which gives rise to the speckle nature of the patterns of Eqn.~\ref{eq:random-speckle}. 
This noise has also been called ``wave interaction noise'', or ``photon excess noise'' by Hanbury Brown and Twiss but was neglected in their signal to noise calculations \cite{HBT_1957, HBT_1957_II}.

We must also consider effects that reduce the visibility of $g^{(2)}$ measurements, such as due to polarization states, temporal incoherence, insufficient sampling of the speckles or energy spread. Since the signal is modulated by the visibility $\beta$, maximizing it is as important as minimizing noise.  Temporal coherence can be considered in terms of modes, whereby photons in a mode are mutually coherent (giving rise to interference) and incoherent to those in separate modes.  The visibility is equal to the inverse of the number of modes, $M = \beta^{-1}$, which can be expressed as~\cite{Goodman} 
\begin{equation}
    M = \frac{2}{1 + \mathcal{P}^2} \left( \frac{1}{T} \int_{-\infty}^\infty \Lambda\left( \frac{\tau}{T} \right) \left| \gamma(\tau) \right|^2 d \tau \right)^{-1} \, ,
\end{equation}
where $0 \leq \mathcal{P} \leq 1$ denotes the degree of polarization, $T$ the measurement time,  $\Lambda(\tau) = 1 - |\tau|$ for $|\tau| \leq 1$ and $0$ otherwise, and $\gamma(\tau)$ is the complex degree of temporal coherence.
For example, a Lorentzian spectrum characteristic of inner-shell fluorescence~\cite{Sorum1987}) with a coherence time $\tau_c$ yields~\cite{Goodman}
\begin{equation}
    M_\text{L} = \frac{2}{1 + \mathcal{P}^2} \left( 
    \frac{\tau_c}{T} + \frac{1}{2} \left( \frac{\tau_c}{T} \right)^2 \left( e^{- 2 (T/\tau_c)} - 1 \right)
    \right)^{-1}\, .
\end{equation}
For measurement times sufficiently greater than the coherence time 
\begin{equation}
    M \approx \frac{2}{1 + \mathcal{P}^2} \frac{T}{\tau_c}\, .
\end{equation}
For x-ray fluorescence and an polarisation-insensitive detector, $\mathcal{P} = 0$, giving a factor $2$ to the number of modes due to the two orthogonal polarisation eigenstates\footnote{The analysis of Inoue \emph{et al.}~\cite{Inoue:gb5094} neglected this factor of 2.}.  

Fig.~\ref{fig_SpeckleDemo} shows four examples of integrated intensities simulated for fluorescence from a small simple cubic crystal and a single emitter located in each of its $15 \times 15 \times 15$ unit cells.  The simulation assumed a lattice constant $a=\SI{5}{\angstrom}$, an emitting wavelength of $\lambda=\SI{2}{\angstrom}$, and a detector placed \SI{50}{\milli\meter} from the sample with pixels of area $\SI{100}{\micro\meter} \times \SI{100}{\micro\meter}$. The four patterns were simulated for different values of the mean intensity per emitter, characterised by the mean detected counts $\mu$ per pixel,  and number of modes $M$.  The high numbers of detected photons per emitter are not physically realisable for single atoms, but the examples serve to illustrate the salient features of such patterns and the noise sources.  When the emitters are very strong, as illustrated with $\mu = 20$ in Figs.~\ref{fig_SpeckleDemo}a and b, the speckle pattern can be discerned, with a speckle width inversely proportional to the object width as discussed in Sec.~\ref{sec:IDI}. The visibility of the speckles is much reduced when the number of modes is increased from 1 to 20 (Fig.~\ref{fig_SpeckleDemo}b), which would be the case for a measurement with a polarisation-insensitive detector and an exposure time $T\approx 10\,\tau_c$. Poisson noise can also be seen in these patterns, but that noise dominates when the number of photons detected per emitter is reduced as in the simulations depicted in Figs.~\ref{fig_SpeckleDemo}c and d. There the effect of the modes is harder to see.

We model the statistics of the photon correlations by considering an object consisting of $N_\text{E}$ emitters that emit monochromatic spherical waves with random relative phases.  The measured energy on a pixelated detector for a single mode is then given by Eqn.~\ref{eq:random-speckle}.  Since the sum of the waves with random phases can be viewed as a random walk in the complex plane, the energy is a random variable with the negative exponential distribution function $P_\text{exp}(x , \mu_0) = \frac{1}{\mu_0} e^{-x / \mu_0}$, where $\mu_0$ is the average energy per pixel and mode~\cite{Goodman:76}. The detected energy for $M$ modes consists of the sum of $M$ such random variables. This generates an Erlang distribution described by
\begin{equation}
    P_\text{Erlang} (x , \mu, M) = \frac{x^{M-1} e^{- \frac{x M}{\mu}}}{\left( \frac{\mu}{M} \right)^M (M-1)!}\, , 
\end{equation}
where $\mu$ is taken to be the average energy per pixel of the measurement, so that $\mu = M \mu_0$. The variance is then given by $\text{Var}_\text{Erlang} = \mu^2 / M$ representing the ``phase noise".

Since photons are detected as countable particles the Erlang-distribution must be combined with a Poisson-distribution, generating a negative binomial (NB) distribution~\cite{Mandel_1959} as
\begin{equation}
    P_\text{NB}(x , \mu, M) = \frac{M^M \mu^x (M+x-1)!}{(M+\mu)^{M+x} x! (M-1)!} \, ,
\end{equation}
where now the random integer variable $x$ describes the number of detected photons at pixels and  $\mu$ is again the mean counts per pixel. The variance is given by 
\begin{equation}
    \text{Var}_\text{NB} = \mu + \frac{\mu^2}{M} \, .
\end{equation}

The first term $\mu$ is equal to the variance for a Poisson distribution (obtained when $M \rightarrow \infty$) and as such represents the contribution of ``Poisson noise", while the $\mu^2/M$ term is equivalent to the variance of the Erlang distribution and so can be considered to be due to ``phase noise" and modes, as can be related back to the examples in Fig.~\ref{fig_SpeckleDemo}. The ``phase noise" is dominant in Fig.~\ref{fig_SpeckleDemo}a, and increasing the number of modes decreases the variance as seen in Fig.~\ref{fig_SpeckleDemo}b. In Fig.~\ref{fig_SpeckleDemo}d, ``Poisson noise" is dominant.
We also recognise that $P_\text{NB}$ becomes the Bose-Einstein-distribution~\cite{MandelWolf} for $M=1$, given by
\begin{equation}
    P_\text{BE}(x, \mu) = \frac{1}{1 + \mu} \left( \frac{\mu}{1 + \mu} \right)^x \, .
\end{equation}

Incoherent diffractive imaging requires the auto-correlation of measured counts as described in Eqn.~\ref{eq:corr-avg2}. The simplest way to perform this correlation is to multiply the counts of two single-pixel detectors. Initially, for the sake of simplicity, we assume that the counts of both detectors are uncorrelated. Their correlation then follows the distribution of the product of two NB-distributed random variables.
The expectation value of this product distribution is $\mu_{\text{NB}\cdot \text{NB}} = \mu^2$, where $\mu$ remains the expectation value of the detected counts: $\mathbb{E}(I) = \mu$. The variance of this product distribution is given by
\begin{equation}\label{eq_VarNBNB}
    \text{Var}_{\text{NB}\cdot \text{NB}} = 
    \frac{1 + 2M}{M^2} \mu^4 + 2 \frac{1+M}{M} \mu^3 + \mu^2 \, .
\end{equation}
Therefore, this relation describes the variance of the correlation of signals measured in two single-pixel detectors (for instance the two telescopes of Hanbury Brown and Twiss \cite{Brown1968}), or for coincidence measurements made between two detectors out of a multiple detector array (as proposed using the Cherenkov Telescope Array \cite{Dravins2013}).
When, on the other hand, measurements are made using a pixelated detector, where the counts in many detector pairs are acquired \emph{simultaneously}, the discrete $\text{AC}(q)$ is given by a sum of such products.
In order to investigate the effect of performing measurements with many pixels, we assume a set of $J$ NB-distributed values $I(j)$, representing the photon counts at the pixels $j$. Further, we assume, for the sake of simplicity, that the angular positions of these pixels evenly spaced along a line of $\vec{k}$ positions (that is, a one-dimensional array). We keep our assumption that $I(j)$ are actually uncorrelated. We further assume periodic boundary conditions: $I(j+J) = I(j)$. The auto-correlation then becomes 
\begin{equation}
    \label{eq:AC}
    \text{AC}(q) = \frac{1}{J} \sum_{j=1}^{J} I(j) I(j-q)\, .
\end{equation}
Each term within this sum follows the product distribution with a variance given by Eqn.~\ref{eq_VarNBNB}. Note that even if there is no correlation between the single multiplicands $I(j)$ (per our assumption), there may still be a covariance between the summands of Eqn.~\ref{eq:AC} . For $I(j)I(j-q)$ and $I(l)I(l-q)$ there is no covariance if $|j-l| \neq q$, but if $j-l = q$ there will be. As an example, consider $q=1$: 
\begin{table}[H]
\begin{tabular}{l | c | c | c | c | c | }
    \cline{2-6}
    & $I_1$ & $I_2$ & $I_3$  & $\dots$ & $I_J$ \\ \cline{2-6}
    & $I_J$ & $I_1$ & $I_2$  & $\dots$ & $I_{J-1}$ \\ \cline{2-6}
   $\frac{1}{J} \sum$ & $I_1 \cdot I_L$ & $I_2 \cdot I_1$ & $I_3 \cdot I_2 $ & $\dots$ & $I_J \cdot I_{J-1}$ \\ 
    \cline{2-6}
\end{tabular}
\end{table}
\noindent
Obviously there is a correlation between the summands $I_2 \cdot I_1$ and $I_3 \cdot I_2$, which must be taken into account to calculate the variance of the auto-correlation, Eqn.~\ref{eq:AC}. Given that the variance of a sum of random variables is the sum of the covariances of all combinations of pairs of those variables, and $\text{Cov}(X,X) = \text{Var}(X)$ (for a random variable $X$), we see that 
\begin{equation}
    \label{eq:VarAC-derivation}
    \begin{aligned}
    \text{Var}_\text{AC} &=\frac{1}{J} \sum_{j,k=1}^J \text{Cov}[I(j) I(j-q), I(k) I(k-q)] \\
    &=\frac{1}{J} \sum_j^J \text{Var}[I(j) I(j-q)] + \\ 
    &\frac{2}{J} \sum_j^J \text{Cov}[I(j) I(j-q), I(j-q) I(j-2q)] \, ,
    \end{aligned}
\end{equation}
since the condition $j-l = q$ appears $J$ times within the auto-correlation sum. Noting that the covariance is given by $\text{Cov}(X,Y) = \mathbb{E}(X\cdot Y) - \mathbb{E}(X) \cdot \mathbb{E}(Y)$ for random variables $X$ and $Y$, we find that the terms in the last sum of Eqn.~\ref{eq:VarAC-derivation} are therefore equal to $\mathbb{E}(I(j)) \cdot \mathbb{E}(I^2(j-q)) \cdot \mathbb{E}(I(j-2q)) - \mu^4$.  The expectation value of the square of a negative binomial distributed variable is $ \mathbb{E}(I^2) = \mu + \mu^2 \left( 1 + 1/M \right)$ and therefore the last sum in Eqn.~\ref{eq:VarAC-derivation} equals $2(\mu^4/M + \mu^3)$.  The first term of the second line of Eqn.~\ref{eq:VarAC-derivation} is equal to $\text{Var}_{\text{NB}\cdot \text{NB}}$, so the complete sum gives the variance of the auto-correlation as
\begin{equation} \label{eq_VarAC}
  \text{Var}_\text{AC} =  \frac{1+4M}{M^2} \mu^4 + 2 \frac{1+2 M}{M}  \mu^3 + \mu^2 \, .
\end{equation}

It is important to note that Eqn.~\ref{eq_VarNBNB} and thus also Eqn.~\ref{eq_VarAC} were derived under the assumption of an absence of correlation between the measured counts, even though this is what would generate the signal that pertains to the structure of the emitting sample.

Further, it should be noted that besides the contribution to $\langle I(\vec{k})\,I(\vec{k}-\vec{q})\rangle_k$ from the desired  $|g^{(1)}(\vec{q})|^2$ arising from the structure, there are contributions to the background that are also correlated when measured simultaneously within a single pattern. While averaging over many patterns smooths the background to a constant value, a $\vec{q}$-dependence of the variance can persist. The background cannot be strictly defined, therefore, as the uncorrelated (or zero covariance) contributions to $\langle \langle I_p(\vec{k})\,I_p(\vec{k}-\vec{q}) \rangle_{\vec{k}} \rangle_p$. 
This is illustrated in Appendix \ref{sec_A_infDet} where the variance is calculated for detectors consisting of two pixels and of many pixels. It is seen there that the correlations contributing to the background change the form of the variance of $g^{(2)}(\vec{q})$ from that indicated by Eqn.~\ref{eq_VarAC}, which can be only considered an approximation. This leads to the effect that increasing the number of pairs of pixels within the same pattern is not equivalent to collecting more patterns. In the following we continue with this approximation and explore the validity of results by comparing with simulations.

\section{Signal to noise ratio} \label{sec_SNR}
We now aim to determine the dependence of the signal to noise ratio (SNR) of IDI on the various kinds of noise discussed above. For this discussion we define our signal as $\text{Sig} = \left|G^{(1)}(\vec{q})\right|^2 = \left| \tilde{S}(\vec{q}) \right|^2$ and the noise as the standard deviation of the background as discussed in the previous section. This signal is proportional to the square of the measured counts, so we can also write $\text{Sig} = \mu^2 \left| \tilde{S}(\vec{q})/\tilde{S}(0) \right|^2$. This is different to CDI, where the signal scales linearly with $\mu$. 

In the following we examine various situations and different kinds of fluorescing samples which require different simulation methods (detailed in Appendices~\ref{sec_A_3DSim} and \ref{sec_A_2DSim}).  The detector arrangements also differ, according to sampling requirements, placing the different cases on quite different scales and making direct comparisons somewhat artificial (for example imaging a crystal versus a single non-periodic object).  We therefore concentrate on separately studying the dependence of the SNR on varying intensity, numbers of modes, and object shape to gain an understanding of how to best design experiments.

Following Eqn.~\ref{eq_VarAC} we express the SNR as
\begin{equation}\label{eq_SNR_Theo}
    \text{SNR} (\vec{q}) = \frac{ \text{Sig} (\mu, \vec{q}) \, \sqrt{C(\vec{q})} \, \sqrt{N_\text{P}}}{M\, \sqrt{ \frac{1+4M}{M^2}  \mu^4 + 2 \frac{1+2 M}{M}  \mu^3 + \mu^2 }} \, .
\end{equation}
where $C(\vec{q})$ is the multiplicity equal to the number of pixel pairs with the same wave-vector difference. The multiplicative factor $1/M = \beta$ accounts for the visibility of $|G^{(1)}|^2$.
One should note that increasing $C(\vec{q})$ does not have the same effect on the SNR as increasing $N_\text{P}$ and this term saturates at some point. Even if we consider a detector covering $4\pi$ with infinite sampling, a single pattern will still suffer from phase-noise, since the assumption of independent photon counts forming the background can not be maintained. A simple, analytic example is discussed in Appendix~\ref{sec_A_infDet}.

As a first example we consider a crystal with $n \times n \times n$ simple cubic unit cells. We assume each unit cell consists of one cluster of single photon emitters that are so close to each other that they are indistinguishable and can be treated as one emitter. The crystal then consists of $N_\text{E} = n^3$ emitters, each isotropically emitting on average $N_\gamma$ photons per mode and pattern. The expected mean counts per detector pixel therefore is $\mu = \Omega N_\text{E} N_\gamma M$, where $4 \pi \Omega$ is the solid angle of a pixel (here, for the sake of simplicity, assumed to be constant over the whole detector). The autocorrelation signal $G^{(2)}(\vec{q}) = \langle \langle I_p(\vec{k})\, I_p(\vec{k}-\vec{q})\rangle_{\vec{k}} \rangle_{p}$ obtained from the measured fluorescence photon counts of the crystal consists of a uniform background with strong peaks at the reciprocal lattice points (Bragg peaks) as shown in Fig.~\ref{fig_Sim3D}. 
The $|G^{(1)}|^2$ map that is extracted from the autocorrelation can be written as
\begin{equation} \label{eq_CrystG1}
    \left| G^{(1)}(\vec{q},N_\text{E}) \right|^2 \propto
    \left| \prod_{j=x,y,z} \left( \frac{1}{\sqrt{2 \pi}} \frac{e^{i \sqrt[3]{N_\text{E}} a q_j}-1}{e^{i a q_j}-1} \right) \right|^2\, ,
\end{equation}
where $a$ is the lattice constant. We then define the signal that is extracted from such a map as the values integrated over Bragg peaks, which in the limit of large cubic crystals is proportional to the number of emitters: 
\begin{equation}
    \label{eq:Bragg-integration}
    \lim_{N_\text{E} \to \infty} \int_{-\frac{\pi}{a}}^{\frac{\pi}{a}} \left| G^{(1)}(\vec{q},N_\text{E}) \right|^2 dq_x\, dq_y\, dq_z = N_\text{E} \, .
\end{equation}
This yields a signal as described by
\begin{equation} \label{eq_CrystSig}
    \text{Sig}_{\text{Crystal}} = N_\text{E}  N_\gamma^2 \Omega^2 M^2 = \frac{ \mu^2}{N_\text{E}} \, .
\end{equation}
and thus, the SNR is 
\begin{equation}\label{eq_SNRCryst_Theory}
    \text{SNR}_{\text{Crystal}} = \frac{ \mu^2 \sqrt{C(\vec{q})}\sqrt{N_\text{P}}}{N_\text{E} M \sqrt{\frac{1+4M}{M^2}  \mu^4 + 2 \frac{1+2 M}{M}  \mu^3 + \mu^2}} \, .
\end{equation}

A word of caution about Eqn.~\ref{eq_SNRCryst_Theory} is warranted since it indicates that pixels of larger solid angle should result in higher SNR. In fact, as the pixel size is increased, the number of modes increases in accordance with a loss of contrast~\cite{Dainty:1976}. 
Since this effect can be treated by an appropriate adjustment of the number of modes we further ignore the ``speckle sampling'' effect in this paper to keep it as simple as possible.

\subsection{SNR as function of mean counts}\label{sec_Mu}
\begin{figure}
   \def\svgwidth{150pt}
   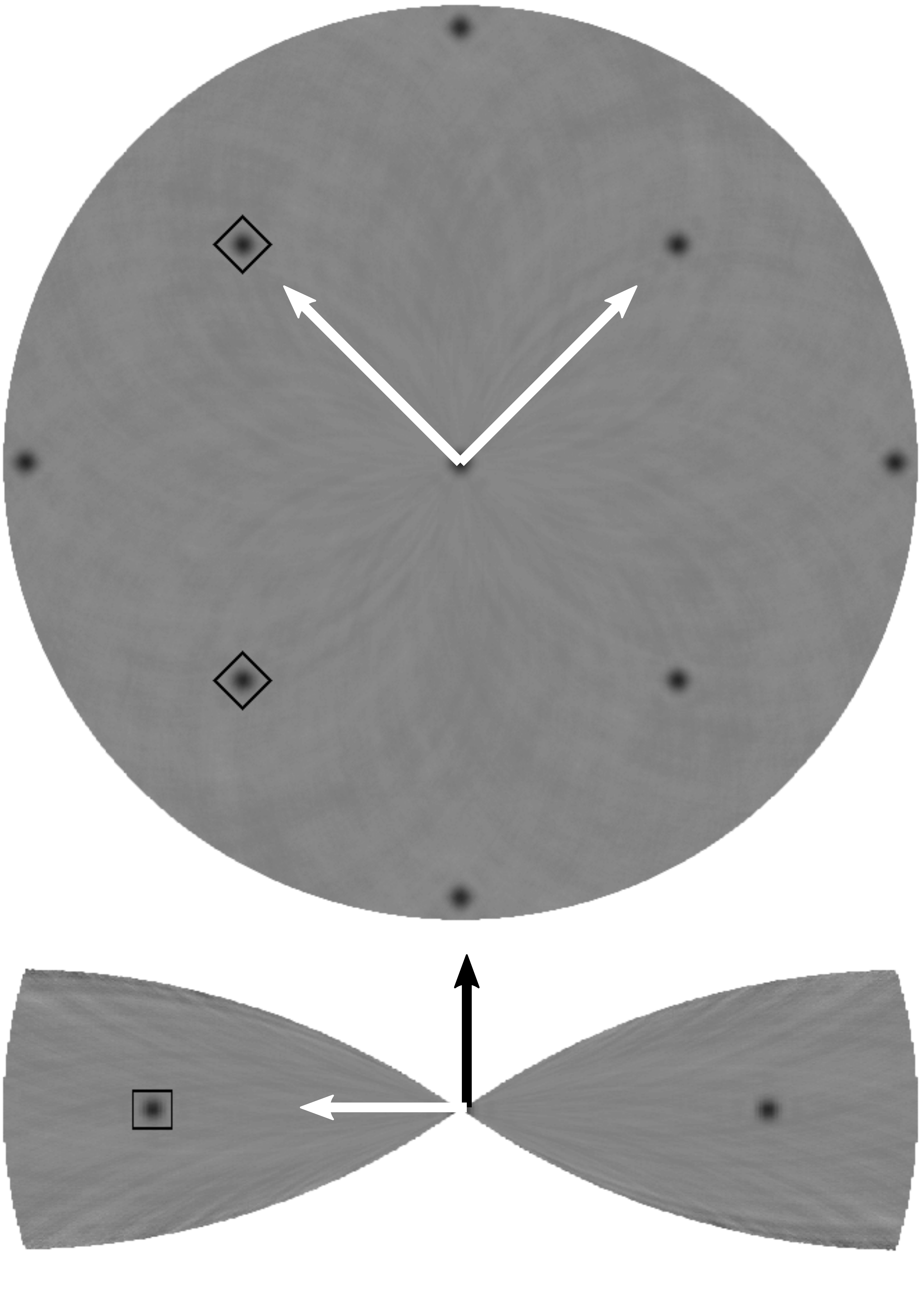
\caption{\label{fig_Sim3D} Slices through $G^{(2)}(\vec{q})$-space from incoherent simulation of a $15 \times 15 \times 15$-crystal. Integration boundaries for the signal are indicated as squares around the peaks.}
\end{figure}

To test the SNR expression in Eqn.~\ref{eq_SNRCryst_Theory}, we first investigate the dependence of the SNR of simulated data on $\mu$, or more precisely on $N_\gamma$ (the number of photons emitted by a cluster of non distinguishable emitter per mode). Details of the simulations are given in Appendix \ref{sec_A_3DSim}, and in all simulations in this paper the object (emitter density) consists only of real and positive values. We performed simulations with three-dimensional crystals from which two slices through $G^{(2)}(\vec{q})$ are shown in Fig.~\ref{fig_Sim3D}. 

\begin{figure}
   \resizebox{0.66\linewidth}{!}{\input{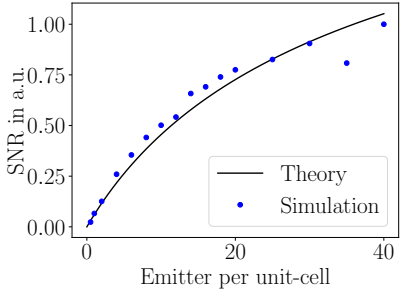}}
\caption{\label{fig_SNR_CrystNE3D} SNR for a simple cubic $15 \times 15 \times 15$ crystal as functions of emitter per unit-cell (located at the same spot within the unit-cell).}
\end{figure}

In Fig.~\ref{fig_SNR_CrystNE3D}, the SNR of the two Bragg peaks highlighted in Fig.~\ref{fig_Sim3D} is plotted as a function of $N_\gamma$, which was changed by adding more emitters to the cluster in each unit-cell, keeping the size of the crystal constant. This is effectively the same as increasing the intensity (emitted number of photons per mode) of each emitter. We observe that the SNR increases with increasing sample intensity but appears to asymptote to a certain value. This is because for a small number of photons, Poisson noise is dominant yielding $\text{SNR} \propto N_\gamma$, whereas for a sufficiently large number of photons per pixel, phase noise becomes important which yields a constant SNR for a fixed number of patterns and modes. This can also be seen from Eqn.~\ref{eq_SNRCryst_Theory} where the low and high-signal limits are $\frac{ \mu \sqrt{N_\text{P}}\sqrt{C(\vec{q})}}{N_\text{E} M}$ and $\frac{ \sqrt{N_\text{P}}\sqrt{C(\vec{q})}}{N_\text{E}\sqrt{1+4M}}$ respectively.

To further investigate the SNR as function of $\mu$ we performed simulations of the detected counts obtained from fluorescing two-dimensional non-crystalline objects. Here we used different objects having less distinguishable signals than the Bragg peaks of crystals. Since the signal could not be readily separated from background and noise, the simulated $G^{(2)}$ was fitted to the ground truth via $G^{(2)} = O + S \cdot |g^{(1)}|^2 + \epsilon$, where the fit parameter $S$ can be interpreted as signal, $O$ as the background, and $\epsilon$ as the noise. A more detailed description of the simulations is given in Appendix \ref{sec_A_2DSim}.
In Fig.~\ref{fig_SNR_Mu2D}, the SNR is plotted for four different objects: two very sparse ones, one crystal-like object and one ``dense" object with spatial frequencies giving a continuously filled Fourier-space. The plots of these SNRs were scaled to asymptote to unity, for comparison. This also demonstrates the limits of the theory with its assumption of uncorrelated values following a negative binomial distribution, applied to the case of correlated values with structural information. As can be seen in the figure, 
the theory fits quite well for objects with sparsely populated $g^{(1)}(\vec{q})$ signals (e.g. for the ``Crystal" object in Fig.~\ref{fig_SNR_Mu2D}), since most of the detected counts are indeed uncorrelated in such objects.
We mention in passing that in the limit of dense and unstructured objects, like the ``Dense" object in Fig.~\ref{fig_Mu2D_Objects}, we were able to fit the $G^{(2)}$-variance, for a single mode, as $\text{Var}_{\text {DenseObj}} = \mu^4 + 6 \mu^3 + \mu^2$.
Because of the strong dependency of the variance of $G^{(2)}$ on the characteristics of the object, as seen by the discrepancies of Eqn.~\ref{eq_SNR_Theo} to the simulations in Fig.~\ref{fig_SNR_Mu2D}, we keep the expression of the variance of Eqn.~\ref{eq_VarAC} for further discussions, but need to keep these limits in mind when fitting this model to the simulated data.

\begin{figure}
  \label{fig_Mu2D}    
  \begin{subfigure}[c]{0.55\linewidth}
     \resizebox{\linewidth}{!}{\input{plots/SNR_Mu_2D.pgf}}
     \subcaption{\label{fig_SNR_Mu2D}}
  \end{subfigure}
  \begin{subfigure}[c]{0.35\linewidth}
     \resizebox{\linewidth}{!}{\includegraphics{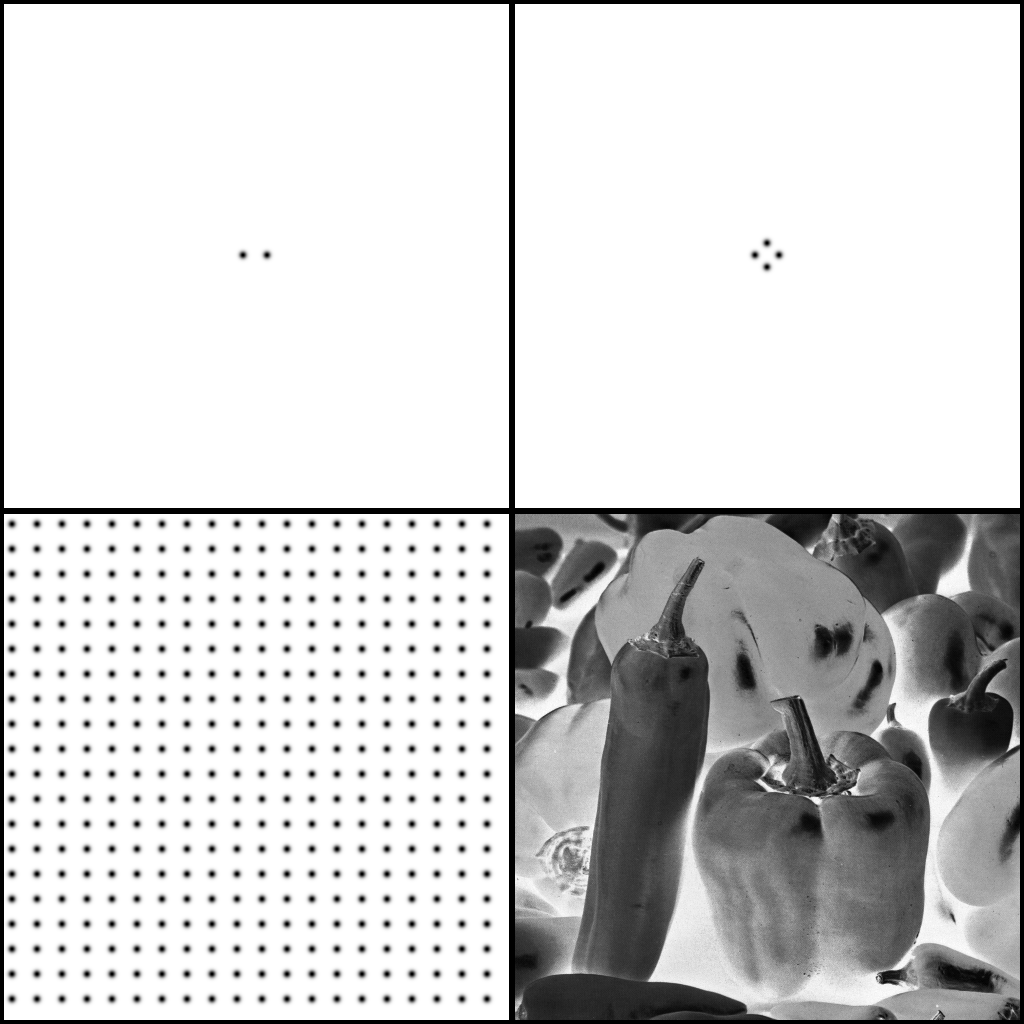}}
     \subcaption{\label{fig_Mu2D_Objects}}
  \end{subfigure}
  \caption{SNR dependence on shape of the object. (a): SNR as function of counts per pixel normalized such that they saturate at the same level. (b): Emitter densities. Top left: ``Double", top right: ``Square", bottom left ``Crystal", bottom right: ``Dense".}
\end{figure}

\subsection{SNR as function of modes}\label{sec_Modes}
Here we discuss the dependence of Eqn.~\ref{eq_SNR_Theo} on the number of modes and make use of the simulations of 3D and 2D objects again. We assume each mode to be of the same mean counts $\mu_0$, so that the total mean counts per pattern is $\mu = M \mu_0$. The SNR then follows the form
\begin{equation}
   \label{eq:SNR-modes}
    \text{SNR} = \frac{ \left| g^{(1)}(\vec{q}) \right|^2 \mu_0^2\,  \sqrt{C(\vec{q})} \sqrt{N_\text{P}}}
    {\sqrt{(1 + 4 M)   \mu_0^4 + (2+4M)  \mu_0^3 + \mu_0^2 }} \, .
\end{equation}

\begin{figure}
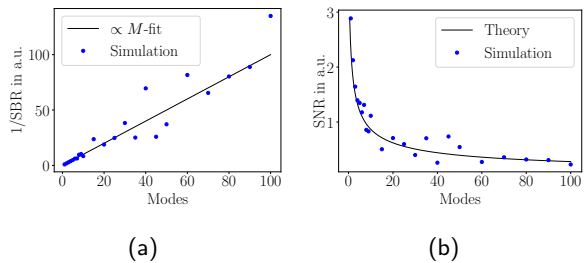


\begin{subfigure}[c]{0.45\linewidth}
   \resizebox{\linewidth}{!}{\input{plots/Modes_Cry3D_SBR_inv.pgf}}
   \subcaption{\label{fig_SBRinv_CrystModes3D}}
\end{subfigure}
\begin{subfigure}[c]{0.45\linewidth}
   \resizebox{\linewidth}{!}{\input{plots/Modes_Cry3D_SNR.pgf}}
   \subcaption{\label{fig_SNR_CrystModes3D}}
\end{subfigure}

\caption{\label{fig_SNR_CrystModes3D_All}Simulation with a $15 \times 15 \times 15$-crystal under variation of modes. Since we set $\mu_0 = 1.35$ to be constant the mean counts per pixel is proportional to $M$. (a): SNR as a function of modes. (b): inverse of the SBR with $\propto M$-fit, to illustrate the $\propto 1/M$ behavior. } 
\end{figure}

As a first example, we consider a simulation of the crystal with $N_\text{E} = 15\times 15\times 15$ unit cells and emitters. This simulation was performed in a similar way to described in the previous section, with a mean counts per mode and pixel of $\mu_0 = 1.35$. 
The reduction of the visibility $\beta = M^{-1}$ with increased modes, according to Eqn.~\ref{eq_g2_beta}, can be seen in the plot of the inverted signal to backgound ratio (SBR) in Fig.~\ref{fig_SBRinv_CrystModes3D}.
The SNR obtained in the simulations is plotted in Fig.~\ref{fig_SNR_CrystModes3D} and found to scale with the number of modes in accordance to the expression in Eqn.~\ref{eq:SNR-modes}.

\begin{figure}
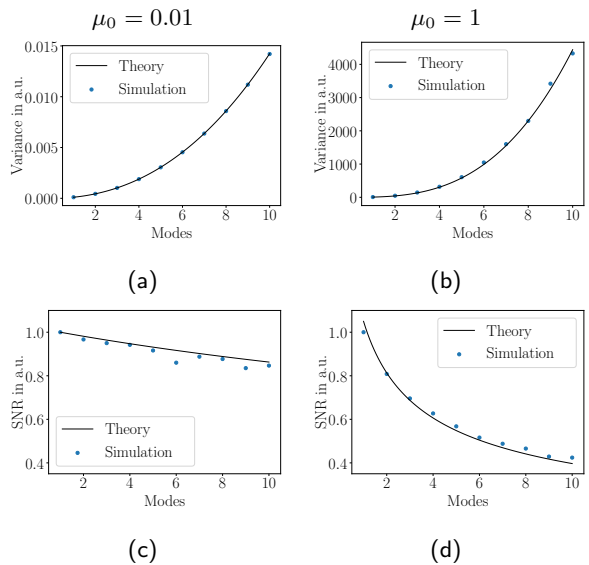

    \begin{tabular}{ c c}
     $\mu_0 = 0.01$ & $\mu_0 = 1$ \\
\begin{subfigure}[c]{0.45\linewidth}
   \resizebox{\linewidth}{!}{\input{plots/VAR_Modes2D_Mu001.pgf}}
   \subcaption{\label{fig_Var_Modes2D_Mu001}}
\end{subfigure}&
\begin{subfigure}[c]{0.45\linewidth}
   \resizebox{\linewidth}{!}{\input{plots/VAR_Modes2D_Mu1.pgf}}
   \subcaption{\label{fig_Var_Modes2D_Mu1}}
\end{subfigure}
\\

\begin{subfigure}[c]{0.45\linewidth}
   \resizebox{\linewidth}{!}{\input{plots/SNR_Modes2D_Mu001.pgf}}
   \subcaption{\label{fig_SNR_Modes2D_Mu001}}
\end{subfigure}&
\begin{subfigure}[c]{0.45\linewidth}
   \resizebox{\linewidth}{!}{\input{plots/SNR_Modes2D_Mu1.pgf}}
   \subcaption{\label{fig_SNR_Modes2D_Mu1}}
\end{subfigure}
\end{tabular}
\caption{\label{fig_SNR_Modes2D_All} $G^{(2)}$-variance and SNR of 'Dense'-object (see Fig. \ref{fig_Mu2D_Objects}) as function of modes. (a,c): Variance and SNR for $\mu_0 = 0.01$. (b,d): $\mu_0 = 1$. Note that the SNR is separately normalized for the two different $\mu_0$.}
\end{figure}

The influence of $\mu_0$ on the mode-dependent SNR was investigated using simulations of the 2D ``Dense" object from Fig.~\ref{fig_Mu2D_Objects}. The variance is plotted in Fig.~\ref{fig_Var_Modes2D_Mu001} and \ref{fig_Var_Modes2D_Mu1} as a function of the number of modes, $M$, for $\mu_0 = 0.01$ and $\mu_0 = 1$. 
 The corresponding plots of the SNR as a function of $M$ are shown in Fig.~\ref{fig_SNR_Modes2D_Mu001} and \ref{fig_SNR_Modes2D_Mu1}, together with the analytic prediction from Eqn.~\ref{eq:SNR-modes}. 

We can see from Fig.~\ref{fig_SNR_Modes2D_All} that the SNR declines much slower with respect to $M$ for $\mu_0 = 0.01$ than for $\mu_0 = 1$.
In the limit of very low $\mu_0$ the dependence of the SNR on $M$ becomes negligible:
\begin{equation}
    \lim_{\mu_0 \to 0} \frac{d}{dM} \text{SNR} (\mu_0,M) = 0\, .
\end{equation}
A negligible dependence of SNR on the number of modes in the limit of low $\mu_0$  (where the contributions of Poisson noise greatly exceeds the phase noise) was already described by Hanbury Brown and Twiss when they stated, that ``...the signal to noise ratio is independent of changes in the optical bandwidth, ..."~\cite{Brown1968}. Similar statements can be found in \cite{Dravins2013, Dravins2015}. Roughly speaking, a slight increase of $\mu$ leads to less Poisson noise, while phase noise is still negligible, and therefore $\mu$ compensates for the weaker visibility caused by a larger number of modes.

In the limit of high intensity per mode, on the other hand, we obtain
\begin{equation}
    \lim_{\mu_0 \to \infty} \text{SNR} (\mu_0, M) \propto \frac{1}{\sqrt{1+4M}}\, ,
\end{equation}
and therefore it is expected that under such circumstances an increase in the number of modes will be significantly detrimental to the SNR.

\subsection{Dependence of SNR on the size and shape of the object} \label{sec_Shape}

\begin{figure}
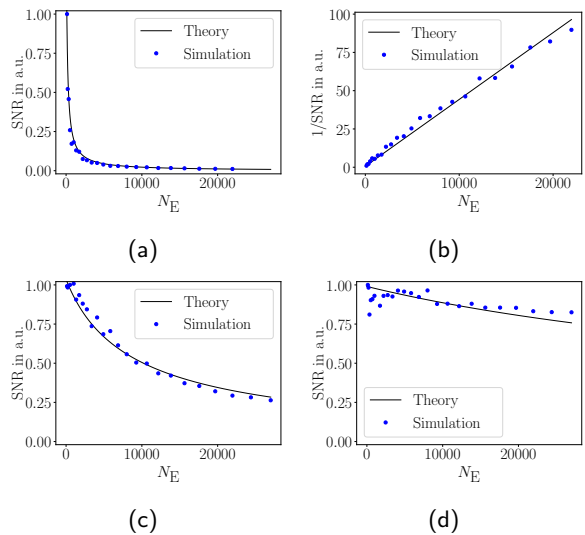

\begin{subfigure}[c]{0.45\linewidth}
   \resizebox{\linewidth}{!}{\input{plots/SNR_CrySz3D_1000.pgf}}
   \subcaption{}
\end{subfigure}
\begin{subfigure}[c]{0.45\linewidth}
   \resizebox{\linewidth}{!}{\input{plots/SNR_CrySz3D_1000inv.pgf}}
   \subcaption{\label{fig_SNR_CrystSize3D_invSNR}}
\end{subfigure} \\

\begin{subfigure}[c]{0.45\linewidth}
   \resizebox{\linewidth}{!}{\input{plots/SNR_CrySz3D_10.pgf}}
   \subcaption{}
\end{subfigure}
\begin{subfigure}[c]{0.45\linewidth}
   \resizebox{\linewidth}{!}{\input{plots/SNR_CrySz3D_1.pgf}}
   \subcaption{}
\end{subfigure}

\caption{\label{fig_SNR_CrystSize3D}SNR as function of crystal size (one emitter per unit cell). (a): $\alpha = 4 \cdot 10^{-3}$ (b): inverse of (a) to demonstrate $\text{SNR} \propto 1/N_\text{E}$ behavior. (c): $\alpha = 4 \cdot 10^{-5}$ (d): $\alpha = 4 \cdot 10^{-6}$. Note that the SNR is separately normalized for the different $\alpha$. }
\end{figure}

In section \ref{sec_Mu} we saw that the shape of the emitting object has a significant influence on the SNR of $G^{(2)}$. Here we return to the 3D crystal with $N_\text{E} = n\times n\times n$ unit cells and emitters, and examine how the SNR scales with the overall size of the crystal. Therefore we define the proportionality constant $\alpha = \Omega\,N_\gamma M$, in an analogous fashion to $\mu_0$ in the previous section, so that $\mu = \alpha N_\text{E}$. With this we rewrite the SNR of Eqn.~\ref{eq_SNRCryst_Theory} as 
\begin{equation}
    \label{eq:SNR-alpha}
    \text{SNR}_\text{Crystal}(N_\text{E},\alpha) = 
    \frac{ \alpha \sqrt{C(\vec{q})} \sqrt{N_\text{P}}}{M \sqrt{  \frac{1+4M}{M^2}\alpha^2 N_\text{E}^2 + \frac{2+4M}{M}\alpha N_\text{E} + 1 }} \, .
\end{equation}
In Fig.~\ref{fig_SNR_CrystSize3D} the SNR is plotted as a function of the crystal size $N_\text{E}$ for three different emitter ``efficiencies" $\alpha$, all for the case of a single mode.  Somewhat unintuitively, bigger crystals give lower SNR.  In the limit of large $\alpha$ the SNR behaves as $1/N_\text{E}$, as indicated in Fig.~\ref{fig_SNR_CrystSize3D_invSNR} where the reciprocal of the SNR is plotted against $N_\text{E}$. However, the SNR becomes less dependent on $N_\text{E}$ and the curve becomes flatter for smaller $\alpha$. This may seem to be an improvement over larger $\alpha$, but for a given crystal size a smaller $\alpha$ gives lower SNR. As discussed in section~\ref{sec_Mu}, a greater $\alpha$ generally leads to a better SNR. However, as mentioned earlier, increasing  $\alpha$ by increasing $\Omega$ alone will reduce contrast and reduce SNR.

In conventional crystallography, which makes use of coherent scattering from the crystal, larger crystals clearly produce higher SNR than small ones.  In that case the SNR is proportional to the square root of the number of photons diffracted per Bragg-peak which by an equivalent analysis to Eqn.~\ref{eq:Bragg-integration} is proportional to $\sqrt{N_\text{E}}$ (assuming a scattering by the emitting atoms).   Equation~\ref{eq:SNR-alpha} and the simulations of Fig.~\ref{fig_SNR_CrystSize3D} show the opposite behaviour in IDI.
Even though we have assumed  perfect conditions (i.e.. $M=1$) in the simulations, in a real experiment there are at least two other factors in favour of choosing smaller crystals. 
The first is that larger crystals lead to smaller speckles, which therefore require smaller pixels or a larger crystal to detector distance, which, for a finite detector, reduces the maximum resolution and results in a decrease in $\mu$.  Secondly,
large crystals can lead to the situation that even for exactly simultaneous emission, the difference of paths to the detector from atoms at the extremes of the crystal can exceed the speed of light times the coherence time, contributing an additional source of modes.


\begin{figure}
\begin{subfigure}[c]{0.45\linewidth}
   \resizebox{\linewidth}{!}{\includegraphics{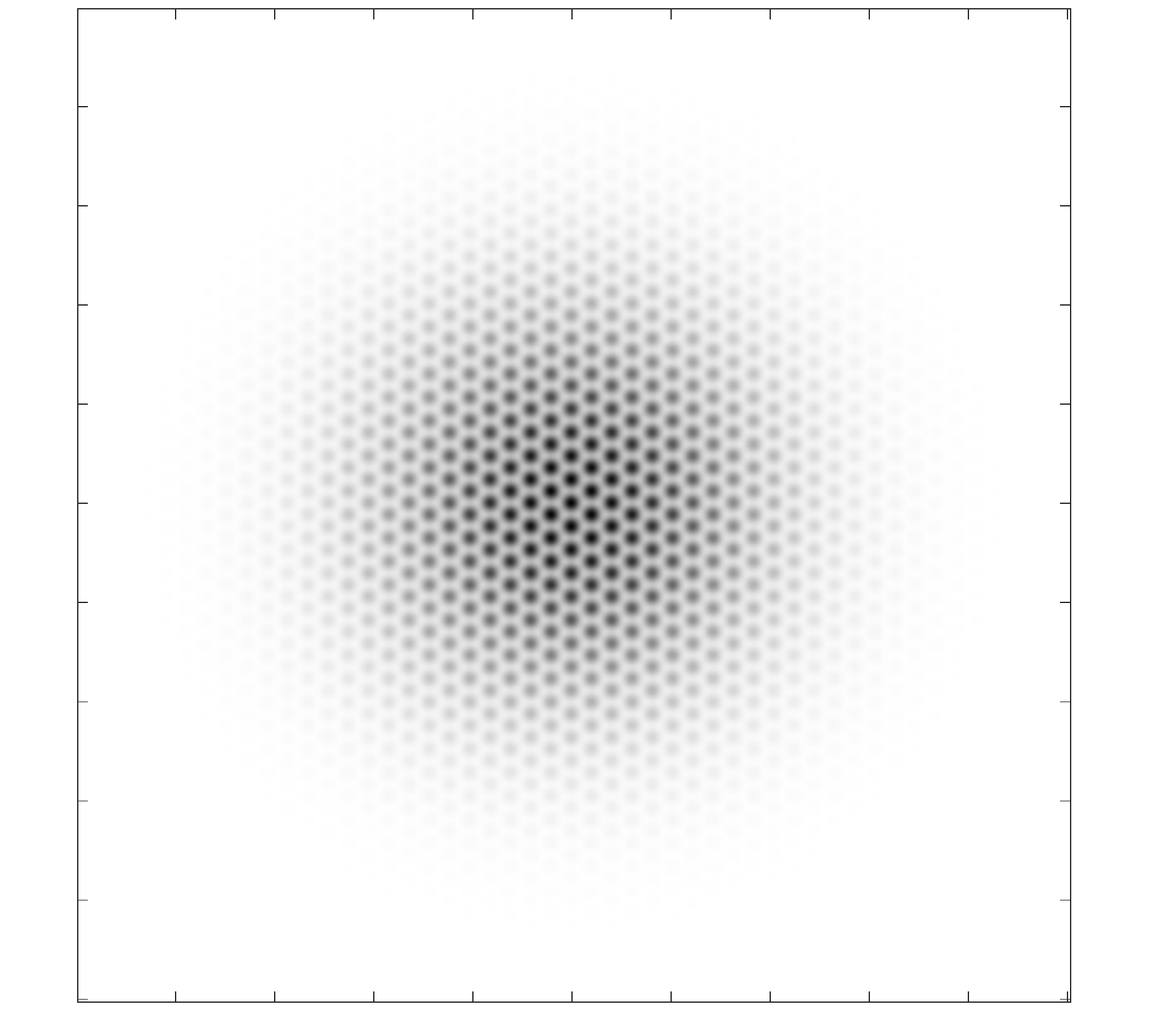}}
   \subcaption{}
\end{subfigure}
\begin{subfigure}[c]{0.45\linewidth}
   \resizebox{\linewidth}{!}{\includegraphics{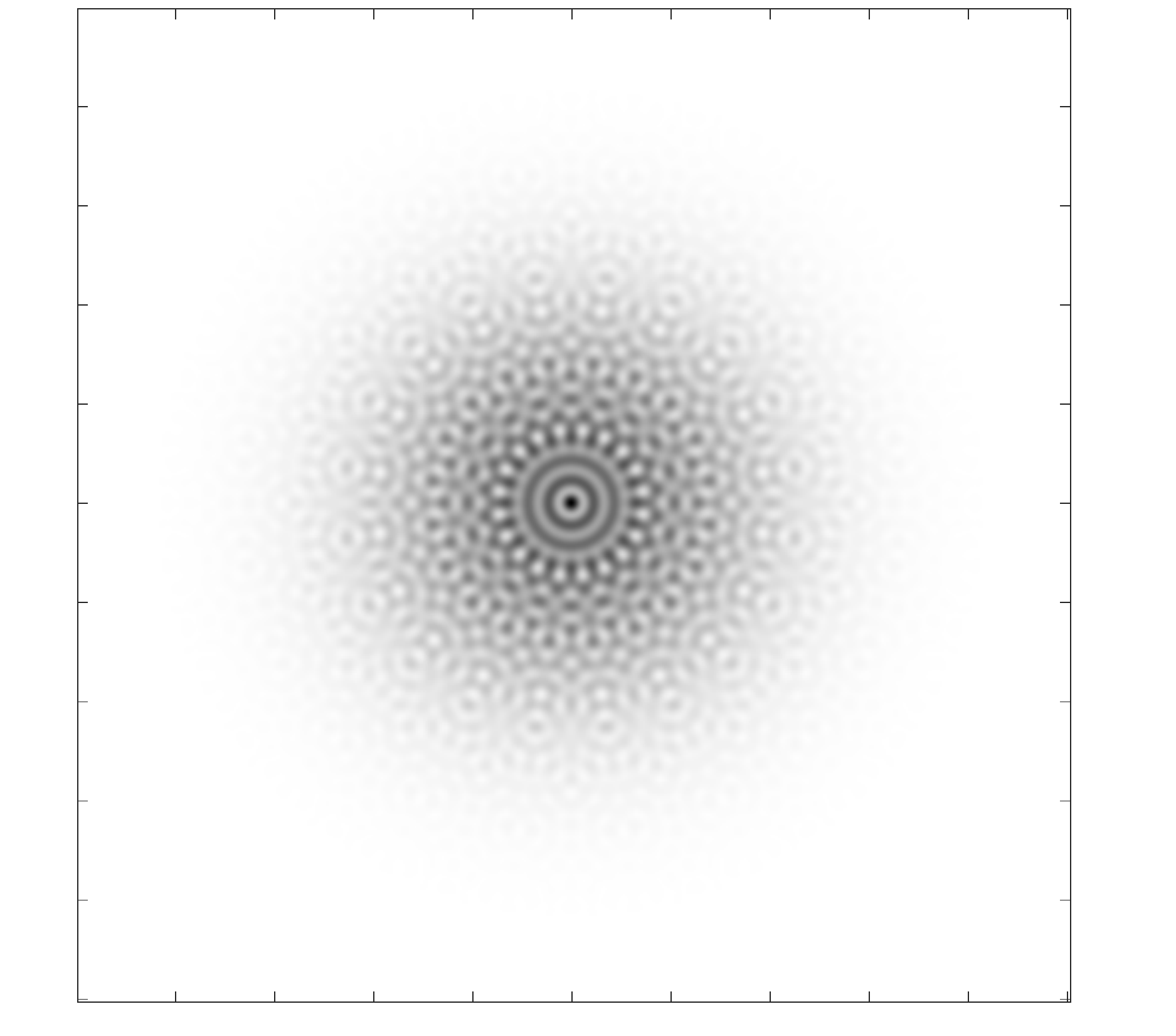}}
   \subcaption{}
\end{subfigure}

\begin{subfigure}[c]{0.45\linewidth}
   \resizebox{\linewidth}{!}{\includegraphics{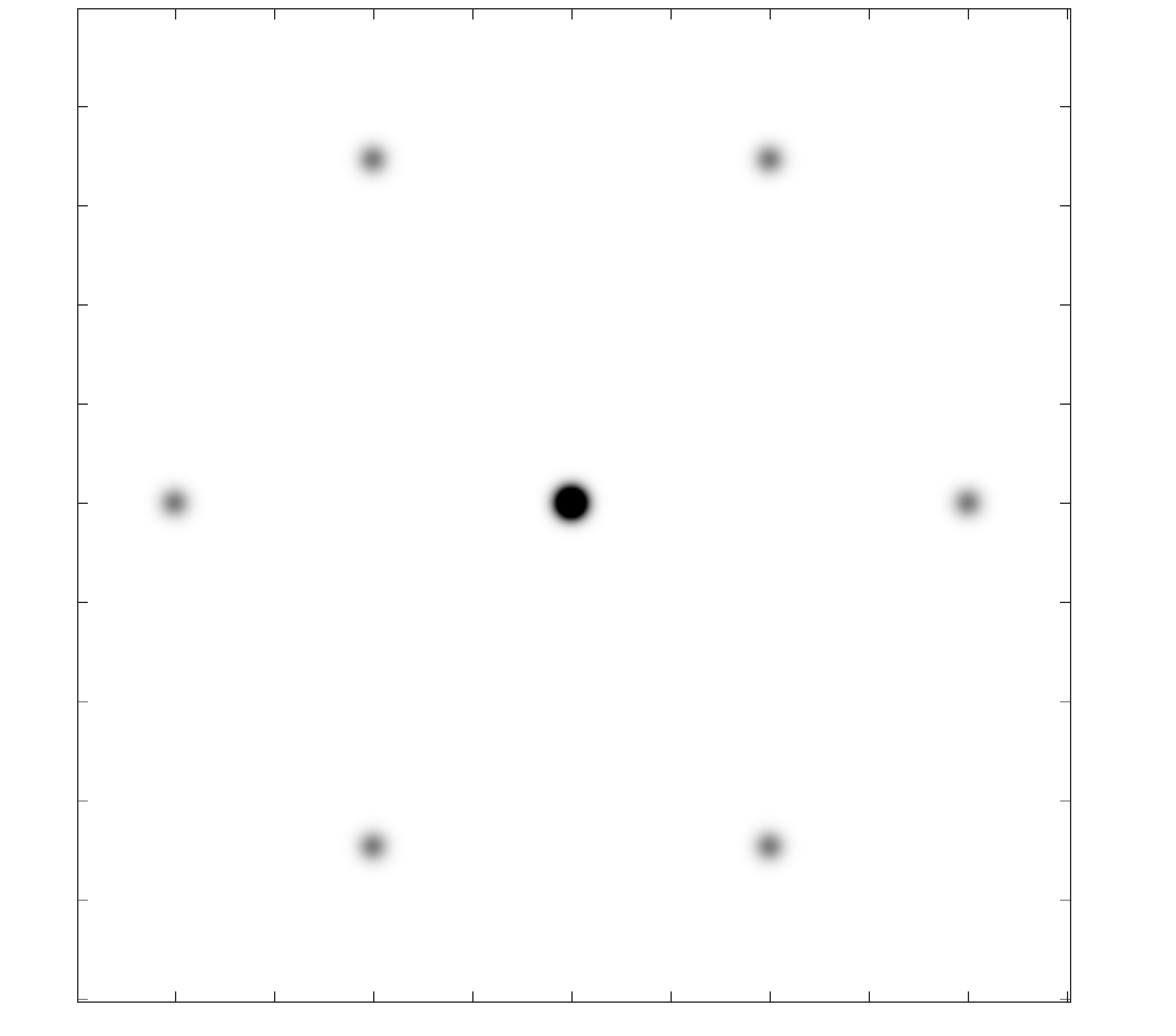}}
   \subcaption{}
\end{subfigure} 
\begin{subfigure}[c]{0.45\linewidth}
   \resizebox{\linewidth}{!}{\includegraphics{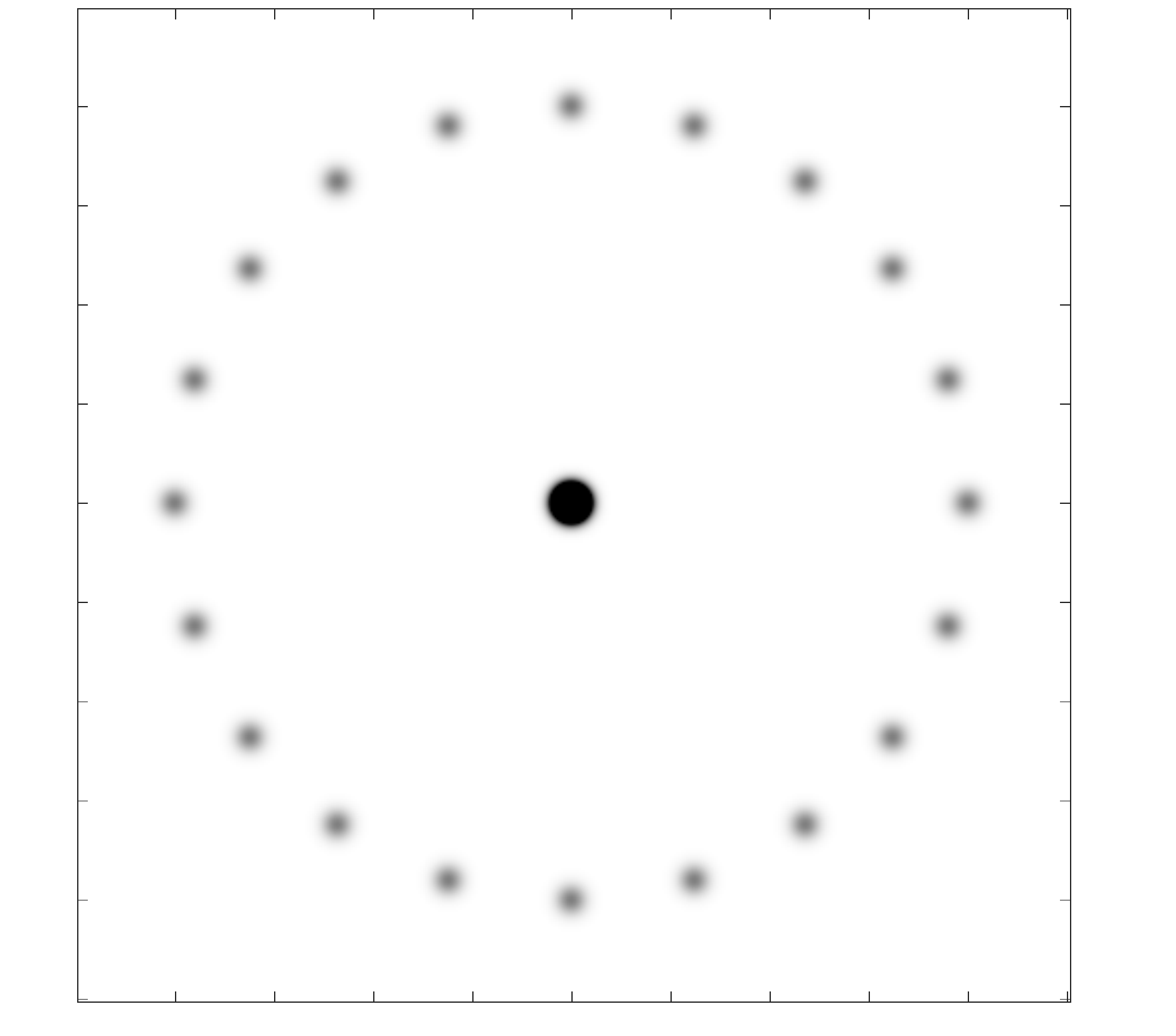}}
   \subcaption{}
\end{subfigure}

\caption{\label{fig_GPS_OBjs} Object with parameterized complexity $c$. (a): Emitter density for $c = 3$ (b): $c=10$ (c,d): are the corresponding $|g^{(1)}|^2$ for (a) and (b). As signal for the analysis the integrated value of one of the outer Gaussians in $|g^{(1)}|^2$-space is used.}
\end{figure}

The reason for the diminishing ability to image larger objects by IDI is due to the fact that as the object gets larger and more complex, the number of intensity-intensity products that do not arise in a correlated signal grows at a greater rate. This is apparent since as $|\tilde{S}(\vec{q})/\tilde{S}(0)|^2=|g^{(1)}(\vec{q})|^2 \leq 1$ for all $\vec{q}$, the background always exceeds, or is at least as large as the the signal for any $q$. Since the distribution of emitters $s(\vec{r})$ is always real and positive, as the object becomes larger $|\tilde{S}(\vec{q})/\tilde{S}(0)|$ generally becomes smaller at any given $q$ as the spectral power is distributed into more ``channels". This is the case if the additional emitters added to a structure (to make it bigger) are resolvable.  Those emitters added close to others (such as considered in the single clusters of emitters in the crystals, above) will tend not to reduce $|g^{(1)}(\vec{q})|$ at $q \ne 0$.  


To investigate the proposition that more complicated objects have lower SNR we carried out simulations of IDI analyses of patterns of non-periodic objects constructed in such a way to give a Fourier spectrum $G^{(1)}(\vec{q})$ consisting of discrete narrow Gaussian-shaped peaks equally spaced in a ring at a particular reciprocal distance $q_1$, as shown in Fig.~\ref{fig_GPS_OBjs}.  The complexity of the object is set by the number of Fourier frequencies that follows from the number of Gaussian peaks, without changing the resolution or overall shape of the object in real space.  The object is parameterised by the number of frequency components in the ring at $q_1$, given by $2\,c$, ensuring a centrosymmetric transform to maintain a real and positive real-space emitter density.
The number of photons per emitter and per pixel is again specified as $\alpha = \Omega N_\gamma M$, and when $\alpha$ is constant the mean counts per pixel is proportional to $c$. We compute the SNR based upon obtaining the signal of the integrated value of $|G^{(1)}(\vec{q})|^2$ of any one of the (non central) peaks. Since the strengths of these peaks do not change with $c$ we assume the signal to be $\text{Sig} \propto \alpha^2$.  With the mean counts per pixel per $\mu = \alpha\,c$, we expect that the SNR scales as 
\begin{equation}
\begin{aligned}
    \label{eq:SNR-C}
    &\text{SNR}(c, \alpha) \propto 
    &\frac{\alpha \, \sqrt{C(\vec{q})} \, \sqrt{N_\text{P}}  }{M \sqrt{\frac{1+4M}{M^2}\alpha^2 c^4 + \frac{2+4M}{M}\alpha c^3 + c^2 }} \, .
\end{aligned}
\end{equation}

\begin{figure}
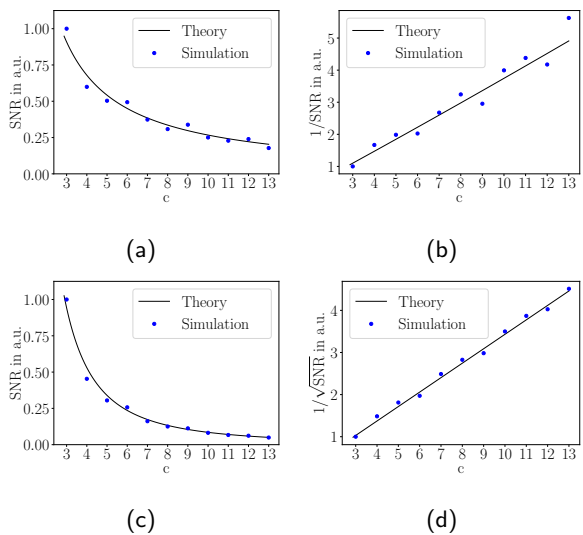

\begin{subfigure}[c]{0.45\linewidth}
   \resizebox{\linewidth}{!}{\input{plots/SNR_GPS_mu0001.pgf}}
   \subcaption{\label{fig_SNR_GPS_lowAlpha}}
\end{subfigure}
\begin{subfigure}[c]{0.45\linewidth}
   \resizebox{\linewidth}{!}{\input{plots/SNR_GPS_mu0001inv.pgf}}
   \subcaption{\label{fig_SNR_GPS_lowAlphainv}}
\end{subfigure}

\begin{subfigure}[c]{0.45\linewidth}
   \resizebox{\linewidth}{!}{\input{plots/SNR_GPS_mu100.pgf}}
   \subcaption{\label{fig_SNR_GPS_highAlpha}}
\end{subfigure} 
\begin{subfigure}[c]{0.45\linewidth}
   \resizebox{\linewidth}{!}{\input{plots/SNR_GPS_mu100inv.pgf}}
   \subcaption{\label{fig_SNR_GPS_highAlphainv}}
\end{subfigure}

\caption{\label{fig_SNR_GPS} SNR as function of ``complexity" parameterized by $c$. (a): Low intensity: $\alpha = 0.001$. (b): inverted SNR ($\alpha = 0.001$) to demonstrate the $1/c$ behavior at low intensities. (c): High intensity $\alpha = 100$. (d): inverted square root of SNR ($\alpha = 100$) to demonstrate the $1/c^2$   behavior at high intensities.}
\end{figure}

The SNR obtained from the simulations based on the parameterised objects is plotted as a function of the complexity parameter $c$ in Fig.~\ref{fig_SNR_GPS}. The case of low intensity, with $\alpha = 0.001$ is shown in Fig.~\ref{fig_SNR_GPS_lowAlpha} and scales as $1/c$, as expected from Eqn.~\ref{eq:SNR-C} which is plotted as the solid line.  Simulations with high photon counts, setting $\alpha=100$, are summarised in Fig.~\ref{fig_SNR_GPS_highAlpha} which show that the SNR scales even more strongly in this case, as $1/c^2$, again in agreement with Eqn.~\ref{eq:SNR-C}.  The simulations support the assertion that the SNR never improves as the object becomes more complex, but instead it most probably becomes worse.

This analysis also implies that in the imaging of stars by intensity interferometry, recovering an image of a binary star (or of a planet transiting a star)~\cite{Dravins2013}, requires overcoming a lower SNR than would be achieved for a single star.

\section{Discussion and conclusions}
\label{sec:summary}
The method of IDI detects x-ray fluorescence that is generated on the exposure of a sample to a short duration of ionising radiation such as a pulse of x-rays from a free-electron laser.  If this generating pulse is of a duration that is comparable to the coherence time of the fluorescence (typically less than \SI{1}{\femto\second}) then the angular distribution of the detected fluorescence will be influenced by the interference of waves originating from the various emitters in the sample. The phases of the emitting waves will be random and different from shot to shot, but the correlations of photon counts measured in a single shot, averaged over many shots, yields a sum of two terms: one that is formed from persisting phase relationships (due to the structure and proportional to the square of the Fourier transform of the structure of emitters) and a term due to correlations of purely random phases.  In the limit of a large number of averages, this second term approaches a constant that is at least as large as the square of the zero-frequency component of the emitting structure.   

An insight gained here from the model and simulations of the IDI measurement, both based upon a classical description of wave interference combined with Poisson photon statistics, is that the optimisation of an IDI experiment, and in particular the requirement of the total number of single-shot patterns to recover the Fourier form factors of the structure of emitters in an object, depends strongly on the size and complexity of the object.  This is apparent from the fact that the background term in the correlation always exceeds the magnitude of all other spatial frequencies of the Fourier spectrum of the object, and as the object becomes more complex the ratio of frequencies to the zero frequency diminishes. Every emitter in the object adds to the background (and therefore the noise) more than it adds to the signal.  In the case of crystals, it was shown in Sec.~\ref{sec_Shape} that an increase in the number of unit cells always decreases the SNR of a particular signal (here, the integrated strength of a Bragg peak). Indeed, for a low number of detected fluorescence counts per emitter, the SNR is inversely proportional to the total number of emitters in the sample. Likewise, the SNR decreases when the number of distinguishable emitters in the object increases, which is the case when the $G^{(2)}(\vec{q})$ function is obtained at a higher resolution (corresponding to higher magnitudes $q$). 

We find also that noise depends not only on Poisson statistics due to photon counting, but also on the structure of the background term.  Poisson statistics are of course familiar to coherent diffraction such as crystallography, where the SNR usually rises in proportion to the square root of the measured counts. The random phases of the emitted waves give rise to a standard deviation in the correlation signal that is proportional to the mean (rather than the square root of the mean). This phase noise was discussed in the context of ``interferometry of intensity fluctuations in light" by 
Hanbury Brown and Twiss~\cite{HBT_1957, HBT_1957_II} (there called ``wave interaction noise"), but not considered by them in further analysis.
We find that phase noise leads to a saturation of the SNR at high intensities, as discussed in section~\ref{sec_Mu}, indicating that higher emission from a given object does not give a proportionally higher SNR. 
Furthermore, as discussed in Sec.~\ref{sec_Modes}, IDI is sensitive to loss of contrast due to mutually incoherent modes (caused, for example, by polarization states, pulse duration relative to the coherence time, finite pixel solid angle, path differences of light emitted from points across the object, and so on). In the limit of high detected counts per mode the SNR is proportional to $1/\sqrt{1+4M}$ for $M$ modes, while for low intensities the influence of modes vanishes due to the dominance of Poisson noise.

Previous analyses of the feasibility of imaging using intensity interferometry have considered only the simple case of intensity measurements using two detectors~\cite{Brown1968} (or multiple detectors where correlations are performed between pairs of detectors (baselines) independently~\cite{Dravins2015}). This is equivalent to using two pixels per exposure (in a larger detector). Here we have examined the case of using a detector with $N_\text{Pix}$ pixels, giving $N_\text{Pix}(N_\text{Pix}-1)/2$ correlations to compute and average for a given reciprocal space vector difference $\vec{q}$, and found that this not equivalent to averaging $N_\text{Pix}(N_\text{Pix}-1)/2$ different shots with a two-pixel detector.  That is, the SNR does not necessarily grow with the square root of the number of correlations that can be performed in an $N_\text{Pix}$-pixel detector, so one cannot make a simple extrapolation from the two-pixel case. This is because the products formed from different combinations of pairs of photon counts exhibit correlations since some pairs share values, as discussed in Sec.~\ref{sec:noise}.  Instead, in the limit of a large number of correlations per shot, the standard deviation of the background ($\sqrt{\text{Var}_\text{AC}}$) can be considerably larger than expected for two detector pixels.  

Our results indicate that IDI may offer utility in structure determination of single molecules at x-ray FELs, using highest possible incident intensities (providing the highest possible number of detected fluorescence photons per atom per pixel per mode), pulse durations comparable to the coherence time, and small object extent (allowing a large solid angle $\Omega$ of pixels). 
The total fluorescence counts from single molecules will be much lower than from macroscopic objects (e.g. the molecule in crystallized form), but the inverse dependence of SNR on the number of emitters shows that the measurement would actually be greatly improved compared with those macroscopic objects. Thus with an optimised detection scheme, IDI could potentially provide element-specific structural information to complement weak coherent scattering~\cite{Neutze:2000}.

\begin{acknowledgments}
This research was supported by the Cluster of Excellence ``Advanced Imaging of Matter" of the Deutsche Forschungsgemeinschaft (DFG) - EXC 2056 - project ID 390715994 and in part through the Maxwell computational resources operated at Deutsches Elektronen-Synchrotron (DESY), Hamburg, Germany. We also acknowledge support from the Initiative and Networking Funds of the Helmholtz Association through ExNet-0002 “Advanced Imaging of Matter.
\end{acknowledgments}


\appendix
\section{IDI simulations of 3D crystals \label{sec_A_3DSim}}
For the IDI simulations of 3D-crystals, we assume a $500\times 500$-pixel detector with a pixel-size of $100 \times$\SI{100}{\square\micro\meter}, placed at a distance of \SI{50}{\milli\metre} to the sample. We consider a cubic crystal sample consisting of simple cubic unit cells with a lattice constant of \SI{5}{\angstrom} and with one emitter per cell.
Each snapshot pattern is simulated by generating a random phase $\phi = [0, 2 \pi)$ for each emitter and mode. The combined scalar wave function arising from the emission of all emitters is calculated for each pixel, making use of the far field approximation and considering a wavelength of \SI{2}{\angstrom}. 
Furthermore, we neglect the quadratic decay of intensity with distance, which is equivalent to the assumption that each pixel covers an equal solid angle. To ensure an accurate representation of the recorded signal, the wave function was evaluated on a grid of nine points that sub-divides each pixel. The continuously-valued intensity for a pixel centered at $\vec{r}_\text{P}$ therefore reads
\begin{equation}
    I_{\text{c}}(\vec{r}_{\text{pix}}) = \frac{1}{9 M} \sum_{m=1}^M \sum^9_{s=1} \left| \sum^{N_\text{E}}_{j=1} e^{i \frac{2 \pi \, \vec{r}_{\text{pix},s} \cdot \vec{r}_{j}}{\lambda |\vec{r}_{\text{P},s}|}} e^{i \phi_{j,m}} \right|^2 \, ,
\end{equation}
where  $\vec{r}_{\text{pix},s}$ are the sampling positions within the pixel at $\vec{r}_{\text{pix}}$ and $M$ is the number of mutually incoherent modes.
The continuously-valued intensity $I_\text{c}$ is then rescaled (according to the fraction of the pixels solid angle $\Omega$, here assumed to be equal for all pixel, and the number of photons per emitter $N_\gamma$, to achieve a certain $\mu$). After that scaling, a Poisson discretization is then applied $I(\vec{r}_\text{pix}) = \text{PoissSampl}(  \frac{\mu}{\langle I_{\text{c}} \rangle} I_{\text{c}}(\vec{r}_\text{pix}))$.

The auto-correlation is calculated as follows
\begin{equation}
\begin{aligned}
    &\text{AC}(\vec{q}) \\ 
    &= \frac{1}{C(\vec{q})} \sum_{j,l=1}^{N_\text{Pix}}  I(\vec{r}_{\text{pix},j})  I(\vec{r}_{\text{pix},l}) \cdot 
    \Pi \left( \frac{2\pi}{\lambda}(\vec{r}_{\text{pix},j}-\vec{r}_{\text{pix},l} ) , \vec{q} \right) \, ,
\end{aligned}
\end{equation}
where $\Pi (\vec{a},\vec{b})$ is defined as a modified top hat function equal to unity if $|a_j - b_j| < \Delta\text{Vox}/2 \, \, | \, \forall_j$ and zero otherwise. $\Delta\text{Vox}$ represents the voxel edge size in a discretized $G^{(2)}$-space. The usage of $\Pi$ therefore represents a nearest-neighbor interpolation of $\vec{q}$. 
If we do not have a spherical $4 \pi$-detector, the number of possible realizations of $\vec{q}$ generally varies. Therefore, we define the function $C(\vec{q})$ as the density of realizations, which reads
\begin{equation}\label{eq_A_Cq}
    C(\vec{q}) = \sum_{j=1}^{N_\text{Pix}} \sum_{l=1}^{N_\text{Pix}}  \Pi \left( \frac{2\pi}{\lambda}(\vec{r}_{\text{P},j}-\vec{r}_{\text{P},l} ) , \vec{q} \right) \, .
\end{equation}
The $G^{(2)}$ is then obtained by averaging $N_\text{P}$ patterns (independent auto-correlations) 
\begin{equation}
    G^{(2)}(\vec{q}) = \frac{1}{N_\text{P}} \sum_p^{N_\text {P}} \text{AC}_{p}(\vec{q}) \, .
\end{equation}

To obtain the variance of $G^{(2)}(\vec{q})$, we perform the whole simulation twice with exactly the same parameters (but with different realisations of the random phases) to obtain $G^{(2)}_1$ and $G^{(2)}_2$. The variance is then estimated by the $C(\vec{q})$-weighted, squared difference of these two auto-correlations:
\begin{equation}
    \text{Var}_\text{Sim3D} = \frac{\sum_j^{N_\text{Vox}} \left( G_1^{(2)}(\vec{q}_j) - G_2^{(2)}(\vec{q}_j) \right)^2 C(\vec{q}_j) }{2 \sum_j^{N_\text{Vox}} C(\vec{q}_j)}\, .
\end{equation}

It should be noted that we have used quite small crystals (starting from $5\times 5 \times 5$ unit-cells) in our simulations. Therefore, the Bragg peaks that arise in $G^{(2)}(\vec{q})$ have non-negligible side maxima that are not easily distinguished from fluctuations in the background. To avoid this we chose to set the integration limits to the positions of the first-order minima $q_{\text{1st min}} = \pm 2\pi/(\sqrt[3]{N_\text{E}} a)$.
Even so, the signal within this integration boundary is only proportional to $N_\text{E}$ in the limit of large crystals.   
Therefore, we calculate peak weighting factors as
\begin{equation}
    \text{PWF}(N_\text{E}) = N_\text{E} \left( \int_{-\frac{2 \pi}{\sqrt[3]{ N_\text{E}} a}}^{\frac{2 \pi}{\sqrt[3]{N_\text{E}} a}} \left| G^{(1)}(\vec{q}, N_\text{E}) \right| dq_x \, dq_y  \, dq_z \right)^{-1} \, ,
\end{equation}
(with $G^{(1)}$ given by Eqn.~\ref{eq_CrystG1})
which are used to scale the integrated Bragg peaks obtained from the simulated $G^{(2)}(\vec{q})$.

\section{IDI simulations of 2D objects\label{sec_A_2DSim}}
IDI simulations of two-dimensional objects were used for the analysis of non crystalline, arbitrary samples. As with the simulations of crystals described in Appendix~\ref{sec_A_3DSim} we assume a detector in the far field, but now the object's emission density is represented by a 2D array of emission values, $\rho(x,y)$, instead of discrete emitters located at arbitrary coordinates. Each emission value of the object is assigned a random phase $\phi_m(x,y) = [0,\, 2\pi)$. The continuously-valued scalar wavefield intensity is then proportional to
\begin{equation}
    I_\text{c}(k_\text{x},k_\text{y}) = \sum_{m=1}^M \left| \text{DFT}^{(2)} \left[ \rho(x,y) e^{i \phi_m(x,y)}  \right]\left(k_\text{x},k_\text{y} \right) \right|^2 \, ,
\end{equation}
making use of the 2D discrete Fourier transform ($\text{DFT}^{(2)}$).
The continuous intensity is represented by a 2D array of the same size as $\rho(x,y)$.
This intensity is then scaled to enforce a given mean pixel intensity $\mu$ and a Poisson discretization is applied ($I = \text{PoissSampl}( \frac{\mu}{\langle I_\text{c} \rangle} I_{\text{c}})$).
The auto-correlation is evaluated as 
\begin{equation}
    \text{AC}(q_\text{x},q_\text{y}) = \text{iDFT}^{(2)} \left[ \left| \text{DFT}^{(2)} \left[ I(x,y) \right] (k_\text{x},k_\text{y}) \right|^2 \right](q_\text{x},q_\text{y}) \, ,
\end{equation}
where $\text{iDFT}^{(2)}$ denotes the 2D inverse discrete Fourier transform.
Contrary to the 3D case with a detector of limited solid angle, here the full two-dimensional $\vec{k}$-space is covered. Therefore,  $C(q_\text{x},q_\text{y}) = N_\text{Pix}$ is constant.
To obtain the signal and the variance the $g^{(1)}$ can be used as the ``ground truth". This is given by
\begin{equation}
    g^{(1)}(q_\text{x},q_\text{y})= \frac{ \left| \text{DFT}^{(2)} \left[ \rho(x,y) \right]\left(q_\text{x},q_\text{y} \right) \right|^2 }
    {\left|  \text{DFT}^{(2)} \left[ \rho(x,y) \right]\left(0,0 \right) \right|^2 }\, .
\end{equation}
The signal and background can now be obtained as fit-parameter ($S$, $B$) with the best fit model
\begin{equation} \label{eq_BgrSig2D_Fit}
    G^{(2)}(q_\text{x},q_\text{y}) = B + S \cdot \left| g^{(1)}(q_\text{x},q_\text{y}) \right|^2 + \epsilon(q_\text{x},q_\text{y}) \, .
\end{equation}
Then the variance is calculated by
\begin{equation}\label{eq_VAR_2D_Fit}
\begin{aligned}
    &\text{Var}_\text{Sim2D} \\
    &= \frac{1}{N_\text{Pix}} \sum_{q_\text{x},q_\text{y}}^{N_\text{Pix}} \left( B + S \cdot \left| g^{(1)}(q_\text{x},q_\text{y}) \right|^2  - G^{(2)}(q_\text{x},q_\text{y}) \right)^2 \, .
\end{aligned}
\end{equation}
It should be noted that for the fitting (Eqn.~\ref{eq_BgrSig2D_Fit}) and the calculation of the variance (Eqn.~\ref{eq_VAR_2D_Fit}) the zero-frequency component ($q_\text{x} = q_\text{y} = 0$) is ignored. This is done because that component follows a different distribution (the squared of a negative binomial distributed value) to that of the ``autocorrelated negative binomial distribution", discussed in section~\ref{sec:noise}.

\section{Examples for the dependence of the variance of $G^{(2)}$ on the detector configuration and correlations within the background term} \label{sec_A_infDet}

In Sec.~\ref{sec:noise} the derivation of the variance of the autocorrelation,  $\text{Var}_\text{AC}(q)$, depends upon the strong assumption that the counts measured at different detector pixels are uncorrelated. Since this assumption may seem quite unsatisfactory, here we illustrate some of the problems one has to face when dropping that assumption. Also, we demonstrate that increasing $N_\text{P}$ and $C(\vec{q})$ do not have the same effect on the SNR in the limit of large values.

We consider a simple one-dimensional arrangement of emitters and further simplify the analysis by adopting the high-intensity limit where Poisson statistics can be neglected (no Poisson noise) and the calculated detected energies are not necessarily discrete. Certainly, the inclusion of Poisson statistics will not make the situation in any way less complicated.

As a first sample we choose two emitters at the positions $r_1 = 0$ and $r_2 = R$. We further assume that the photon signals are measured with two independent detectors (or two detector pixels) at the positions $k_1 = 0$ and $k_2 = q$.
The correlation sum can then be written as
\begin{equation}
\begin{aligned}\label{eq_II_2E2D}
    &I(k_1=0) \cdot I(k_2=q)  \\ 
    &= \sum_{j,j',l,l'}^2 e^{i( k_1 (r_j - r_l) + \phi_j - \phi_l)} e^{-i( k_1 (r_{j'} - r_{l'}) + \phi_{j'} - \phi_{l'})}  \\ 
    &= 2 + 2\cos{(q\cdot r + \phi_1 - \phi_2)} \, ,
\end{aligned}
\end{equation}
and by averaging over many realisations of phases we obtain $G^{(2)}$ as
\begin{equation}
    G^{(2)}(q) = 4 + 2 \cos{(q \cdot R)}\,.
\end{equation}
This expression might seem to contradict Eqn.~\ref{eq_g2} or Eqn.~\ref{eq_g2_SPE} since we now have $g^{(2)} = 1 - 1/N_\text{E} + |g^{(1)}|^2$. This is because Eqn.~\ref{eq_II_2E2D} is not a perfect representation of TLS emitters since that equation only allows up to two photons to be emitted per emitter (and only in the case $j=j'=l=l'$). However, the expression also does not correspond to the pure SPE case. These differences are not relevant to the SNR discussion in the main part of this paper, since they vanish in the limit of large $N_\text{E}$.
However, because of the small number of emitters in this example we are able to calculate the variance of $G^{(2)}$ analytically by integrating over all possible combinations of the random phases. Generally, for objects with $N_\text{E}$-emitters (and therefore $N_\text{E}$ random phases $\phi$) the variance reads
\begin{equation}
\begin{aligned} \label{eq_AnaVar_cont}
    &\text{Var}(q) = \\ 
    &\frac{1}{(2\pi)^{N_\text{E}}} \int_0^{2\pi} \cdots \int_0^{2\pi}  \left(G^{(2)}(q) - I(0,{\phi})\cdot I(q,\{\phi\}) \right)^2 \,\, d\{\phi \} \, .
\end{aligned}
\end{equation}
For our two-emitter object, we therefore obtain the variance as $\text{Var} = 18 + 16 \cos{(q \cdot R)}$. 
If we alter the situation to use more than two independent detectors---say, an infinite number of detector pixels in this thought experiment---covering the full relevant area from $q=0$ to $q=2 \pi / R$, we can write the correlation as
\begin{equation}
\begin{aligned} \label{eq_2E_infDet}
    &\frac{R}{2\pi} \int_0^{\frac{2 \pi}{R}}  I(k,\phi_1,\phi_2) \cdot I(k+q,\phi_1,\phi_2)  \,\,d k \\ 
    &= 4 + 2\cos{(q\cdot R)} \, \,| \, \, \forall_{\phi_1, \phi_2}\, .
\end{aligned}
\end{equation}
Here the variance is obviously zero. That may not seem so surprising, since, under the assumption of uncorrelated photon counts, more detector pixels could be seen as equivalent to more patterns.

To further examine this we alter the sample to three emitters ($r_1 = 0$, $r_2 = R/2$ and $r_3 = R$). Then, the $G^{(2)}$ is given by
\begin{equation} \label{eq_3Em_G2}
    G^{(2)}(q) = 9 + 4 \cos{\left( \frac{q \cdot R}{2} \right) } + 2 \cos{(q \cdot R)} \, .
\end{equation}
The variance is calculated with Eqn.~\ref{eq_AnaVar_cont} (assuming two detectors) and reads
\begin{equation}
\begin{aligned}
    &\text{Var}_\text{2 det}(q) = \\ 
    &142 + 176 \cos{\left( \frac{q \cdot R}{2} \right)} + 88 \cos{(q \cdot R)} + 8 \cos{ \left( \frac{3}{2} q \cdot R  \right)}\, .
\end{aligned}
\end{equation}
We also calculate the case for the ``infinite'' detector in analogy to Eqn.~\ref{eq_2E_infDet} and obtain
\begin{equation}
\begin{aligned} \label{eq_3E_infDet}
&\frac{R}{4\pi} \int_0^{\frac{4 \pi}{R}}  I(k,\phi_1,\phi_2) \cdot I(k+q,\phi_1,\phi_2)  \,\,d k \\ 
&= 9 + 4 \cos{ \left( \frac{ q \cdot R}{2}  \right)} + 2 \cos{(q \cdot R)} \\
&+ 2 \cos{ \left( \frac{ q \cdot R}{2} - \phi_1 + 2 \phi_2 - \phi_3  \right)} \\ 
&+ 2 \cos{ \left( \frac{ q \cdot R}{2} + \phi_1 - 2 \phi_2 + \phi_3  \right)} \, .
\end{aligned}
\end{equation}
The different integration boundary to that of Eqn.~\ref{eq_2E_infDet} is required to sample the full diffraction information. Since the smallest distance in the three-emitter setting is half of the distance between the two emitters in the previous example the integration (Eqn.~\ref{eq_2E_infDet}) in the $q$-space must be doubled.

\begin{figure}
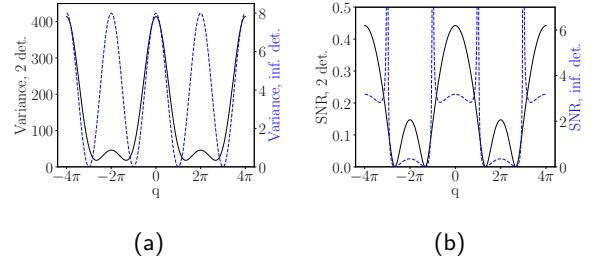

\begin{subfigure}[c]{0.45\linewidth}
   \resizebox{\linewidth}{!}{\input{plots/AnaIDI_VAR_3E.pgf}}
   \subcaption{\label{fig_AnaIDI_Var3E}}
\end{subfigure}
\begin{subfigure}[c]{0.45\linewidth}
   \resizebox{\linewidth}{!}{\input{plots/AnaIDI_SNR_3E.pgf}}
   \subcaption{\label{fig_AnaIDI_SNR3E}}
\end{subfigure}

\caption{\label{fig_AnaIDI_3E} One-dimensional object consisting of three incoherent emitters (with a distance of 0.5). (a) Variance as a function of $q$ for two detectors separated by $q$ (solid black line) and for a 1D-detector of infinite sampling (dashed blue line), covering the full $q$ space. Note that there is not only a difference in scaling but also in the form of the variance. (b) SNR as a function of $q$ for the same object and detector configuration as in (a). Note that for the ``infinite" detector, the SNR maxima are not at the points of maximal signal (see Eqn.~\ref{eq_3Em_G2}).}
\end{figure}

We see that, as opposed to the case in Eqn.~\ref{eq_2E_infDet}, the single-pattern measurement is dependent on the random phases. Therefore, averaging over pixels within a single pattern is not be equivalent averaging over more realisations of patterns with fewer pixels. In other words, the effect of the $C(\vec{q})$ on the SNR is limited.
After averaging over the random phases in Eqn.~\ref{eq_3E_infDet} we obtain the same result as in Eqn.~\ref{eq_3Em_G2}, as expected. 
When calculating the variance using Eqn.~\ref{eq_3E_infDet} we obtain
\begin{equation}
\begin{aligned}
    &\text{Var}_\text{inf det}(q) = 4 + 4 \cos(q \cdot R)\, .
\end{aligned}
\end{equation}
This variance differs from that with only two independent detectors not only in terms of scaling, but also in terms of the its dependence on $q$, as seen in Fig.~\ref{fig_AnaIDI_Var3E}. These differences originate from the fact that the intensity measurements within one pattern are not only correlated due to the emission structure of the object, but also because the terms that form the background are correlated. 
This also leads to the situation that the maxima of the SNR (see Fig.~\ref{fig_AnaIDI_SNR3E}) are not necessarily at the same $q$-positions as the maxima of the $G^{(2)}$.

\newpage

\bibliography{apssamp}

\end{document}